\documentclass[12pt,a4paper]{article}
\usepackage{jheppub}

\newcommand{\ba}{\begin{aligned}}
\newcommand{\ea}{\end{aligned}}
\newcommand{\bej}[1]{ \begin{equation}\label{#1} }
\newcommand{\eej}{\end{equation}}
\newcommand{\beaj}[1]{\begin{eqnarray}\label{#1} }
\newcommand{\eeaj}{\end{eqnarray}}

\def\ZZZ{{\hskip-3pt\hbox{ Z\kern-1.6mm Z}}}
\def\zzz{{\hskip-3pt\hbox{ z\kern-1mm z}}}

\def\a {\alpha}
\def\l {\lambda}

\newcommand{\be}{\begin{equation}}
\newcommand{\ee}{\end{equation}}
\newcommand{\ben}{\begin{eqnarray}\displaystyle}
\newcommand{\een}{\end{eqnarray}}

\def\one{{\hbox{ 1\kern-.8mm l}}}
\def\zero{{\hbox{ 0\kern-1.5mm 0}}}

\title{Generating string solutions in BTZ}

\author{Justin R. David$^a$, Chrysostomos Kalousios$^{b,\,c}$ and }
\author{Abhishake Sadhukhan$^a$}

\affiliation{$^a$Centre for High Energy Physics,
Indian Institute of Science,\\ C.V. Raman Avenue, Bangalore 560012, India}

\affiliation{$^b$Institut f\"ur Physik der Humboldt-Universit\"at zu Berlin,\\
 Newtonstra{\ss}e 15, D-12489 Berlin, Germany}

\affiliation{$^c$ICTP South American Institute for Fundamental Research\\
Instituto de F\'\i sica Te\'orica, UNESP-Universidade Estadual Paulista\\
R. Dr. Bento T. Ferraz 271 - Bl. II, 01140-070, S\~ao Paulo, SP, Brasil}

\emailAdd{justin@cts.iisc.ernet.in}
\emailAdd{ckalousi@ift.unesp.br}
\emailAdd{abhishake@cts.iisc.ernet.in}

\preprint{
\small{\vbox{\hbox{HU-EP-12/44} \hbox{ICTP-SAIFR/2012-008}}}
}

\abstract{
Integrability of classical strings  in the BTZ black hole
enables the construction and study of classical string propagation
in this  background.
We first apply the dressing method to obtain classical string solutions
in the BTZ black hole.
We  dress  time like geodesics in the BTZ black hole and obtain
open string solutions  which are pinned
on the boundary at a single point and whose end points move on time like geodesics.
These strings upon regularising their charge and spins have
a dispersion relation similar to that of giant magnons.
We then dress space like geodesics
which start and end on the boundary  of the BTZ black hole
and obtain minimal  surfaces  which can penetrate the horizon of the
black hole while being pinned at the boundary.
Finally we  embed the giant gluon solutions in the BTZ background
in two different ways. They can be embedded as a spiral
which contracts and expands touching the horizon
or a spike which originates from the boundary
and touches the horizon.
}

\begin{document}
\maketitle
\section{Introduction}

Classical solutions of strings moving in $AdS_5\times S^5 $ backgrounds have played an
important role in the AdS/CFT correspondence.
Spinning strings in $AdS$ or the sphere have been established as excitations dual to operators
with large spins or  R-charges respectively \cite{Gubser:2002tv}. These solutions
were crucial in discovering the role of integrability and verifying many of its
predictions in the AdS/CFT correspondence, see \cite{Tseytlin:2010jv} for a review
and a comprehensive list of references.
Conversely integrability of strings in $AdS$ was used to generate many new and useful
solutions dual to single trace operators as well as Wilson loops in the field theory dual
\cite{Spradlin:2006wk,Kalousios:2006xy,Jevicki:2007pk,Jevicki:2007aa}.

The $AdS_3\times S^3$ dual pair is another example where the role of integrability is
beginning to be investigated
\cite{David:2008yk,Babichenko:2009dk,David:2010yg,OhlssonSax:2011ms,Sundin:2012gc,Sax:2012jv,Ahn:2012hw,Sax}.
One of the most interesting aspects of this dual pair
is that  classical string propagation in the background of $AdS_3$  black holes
is integrable unlike the higher dimensional examples \cite{David:2011iy}.
Integrability can be used
to classify and generate new classical string solutions in the BTZ background.
This in turn will shed light on aspects of black hole physics which can be probed
by extended objects. In this paper we apply  the dressing method
introduced in the AdS/CFT context by \cite{Spradlin:2006wk},
  to generate  and study new solutions of classical strings
in the BTZ background.

Trajectories of point like objects described by
geodesics are the canonical probes of the causal structure of the black hole.
 Space like geodesics in black hole backgrounds are  used to  obtain semi-classical limits of two point
 correlators in the boundary
 \cite{Louko:2000tp,Balasubramanian:1999zv,Kraus:2002iv,Fidkowski:2003nf,Festuccia:2005pi}
and
 to study entanglement entropy of the boundary theory for 3 dimensional backgrounds
\cite{Nishioka:2009un}.
Studying the behaviour of extended objects gives access to new phenomena near black hole
horizons. For instance general arguments indicate that strings are expected to
spread and become tensionless near black hole horizons due to quantum fluctuations
\cite{Susskind:1993aa,Mezhlumian:1994pe,Larsen:1998sh}.
Furthermore minimal surfaces whose boundary are pinned at asymptotic infinity
are dual to Wilson/Polyakov  loops \cite{Maldacena:1998im,Rey:1998bq}  and
are useful probes of the transition from thermal AdS to a  black hole in AdS
\cite{Rey:1998bq,Brandhuber:1998bs}
They are also used to evaluate entanglement entropy \cite{Nishioka:2009un}.
See \cite{Hubeny:2012ry} for a nice review and also for some interesting properties of these
minimal surfaces.
Finally studying spinning strings in the background of black holes in AdS provides
clues of the spectrum of the excitations in the dual thermal  CFT analogous  to the
information provided by the spectrum of quasi-normal modes of fields  in the
black hole background.

The simplest kind of classical string solutions are those which are circular and which
wind around the horizon and which then eventually fall into the horizon.
Such solutions in the context of the BTZ black holes were studied in \cite{David:2010yg}.
They were classified in terms of the finite gap solutions of the
BTZ sigma model.   However since the BTZ sigma model is integrable
it is possible to apply the dressing method
to construct more general classical solutions given a seed solution.
In this paper we show how the dressing method developed for
the $SU(1, 1) $ principal chiral model \cite{Jevicki:2007pk,Jevicki:2007aa}
can be used to generate
classical string  solutions for the sigma model on BTZ$\times S^1$.
One of the by-products of this study is the proof that  the dressing
method preserves the Virasoro constraints of the seed solution
\footnote{The authors are not aware of a proof that
the dressing method in general preserves the Virasoro constraints in the existing
literature.}.
This method of generating solutions for the  sigma model
is different from that obtained by the spectral flow
method used to obtain long strings in the $SL(2, R)$ WZW model in \cite{Maldacena:2000hw}.
The Wess-Zumino term which allowed for the possibility of obtaining
new solutions using the spectral flow in the $SL(2, R)$ WZW
model is not present in the sigma model on BTZ$\times S^1$  considered in this paper.

We first apply the dressing method on time like geodesics to obtain
classical string configurations.
We obtain open string configurations which are pinned at the boundary but
cross the horizon. The end points of these strings move on time like geodesics
which have the same constants of motion as the seed geodesic but
different initial condition. After a suitable regularization of the
energy $E$  and spin $S$  of these solutions which involves subtracting
the energy and spin density of seed geodesic we find that  their dispersion relation
is by the form
\begin{equation}\label{disp}
 E- S = \kappa |\sin\theta|,
\end{equation}
where $\kappa$ is a function of the background and $\theta$ is the
phase of the dressing parameter. Thus the dispersion relation
resembles that of the giant magnons found in \cite{Hofman:2006xt}.
We next examine the minimal surfaces obtained by dressing
space like geodesics. We show that it is possible to obtain closed strings, they
also have a dispersion relation given in (\ref{disp}).  These surfaces are pinned at two
points on the boundary. These points  move on time like trajectories at the boundary.

We finally examine the embedding of the well studied giant gluon solutions
of \cite{Kruczenski:2002fb,Alday:2007hr}
in the BTZ background.
These are Euclidean worldsheet solutions.
We examine two possible embeddings: In one case these solutions have
vanishing energy and spin.  The configuration
is a spiral which originates from the boundary, contracts and touches the horizon and then
expands back to the boundary. In the second solution the embedding  solutions
has a  dispersion relation given by
\begin{equation}
 E+ S = \kappa \log S.
\end{equation}
These are spinning spikes which originate from the boundary and touch the horizon.

The organization of the paper is as follows.
In the next section we adapt the dressing  method developed for the $SU(1,1)$ sigma model
to the BTZ background and show that the Virasoro constraints are preserved by the dressing.
In section 3 we dress time like geodesics and discuss the properties of the
solutions obtained. In section 4 we repeat this analysis for the case of space like
geodesics. In section 5 we embed the giant gluon solutions in the BTZ background and
examine its properties. Section 6 contains our conclusions. Appendix A contains the discussion of the general method to obtain multi-dressed solutions
in the BTZ background.

\section{The BTZ dressing method}

To apply the dressing method to classical solutions in the BTZ background, we first review the
construction of the BTZ black hole as an orbifold of the $AdS_3$ hyperboloid.
Consider the hyperboloid given by
\begin{equation}\label{hyperb}
 -u^2 -v^2 + x^2 + y^2 =-1.
\end{equation}
We then parameterize the hyperboloid as
\be\ba\label{rel1}
 u+x &= \cosh \gamma\, e^{\tilde \phi} , \qquad &u -x  &= \cosh \gamma \, e^{-\tilde\phi},\\
 y+v &= \sinh\gamma\, e^{\tilde t },  \qquad  &y-v &= \sinh\gamma\, e^{-\tilde t}.
\ea\ee
The induced metric on the hyperboloid is given by
\begin{equation}
 ds^2 = d\gamma^2 + \cosh^2 \gamma \, d\tilde\phi^2 - \sinh^2 \gamma \, d\tilde t^2.
\end{equation}
The BTZ black hole is then obtained by the identification
\cite{Banados:1992wn,Banados:1992gq}
\begin{equation}\label{ident}
 \tilde t \sim \tilde t - 2\pi r_-, \qquad \tilde \phi \sim \tilde \phi + 2\pi r_+,
\end{equation}
where $r_+, r_-$ are the inner and outer radii of the BTZ black hole.
The relationship between these coordinates and the conventional radial, time and angular
coordinates are  given by
\begin{equation}\label{rel2}
 \tanh^2 \gamma = \frac{r^2 - r_+^2}{ r^2 - r_-^2}, \qquad
\tilde t = r_+ t - r_- \phi, \qquad \tilde \phi = - r_-t + r_+ \phi.
\end{equation}
The identifications given in (\ref{ident}) ensure that the angular variable $\phi$ has the
required periodicity of $2\pi$.
Note that this parametrization is suitable for $r>r_+$. Suitable parameterizations exist for all the
regions of the black hole. In this paper we will be focussing on the region outside the horizon.
For the BTZ background, it is natural to think of the hyperboloid given in (\ref{hyperb})
as an $SL(2, R)$ group manifold. The group element is given by
\begin{equation} \label{sl2g}
 g  = \left( \begin{array}{cc}
              u + x & y + v \\
y-v & u-x
             \end{array}\right).
\end{equation}
Then the identification given in (\ref{ident}) can be written as
\be \label{ident1}
 g\sim  \tilde A g A, \quad
\tilde A = \left(  \begin{array}{cc}
                    e^{(r_+ - r_-)\pi} & 0 \\
0 & e^{ -( r_+ - r_-) \pi }
                   \end{array}
\right),  \quad
A = \left(  \begin{array}{cc}
                    e^{(r_+ -+r_-)\pi} & 0 \\
0 & e^{ -( r_+  +  r_-) \pi }
                   \end{array}
\right).
\ee
These embedding coordinates are related to the conventional coordinates $r$, $t$ and $\phi$
by the following equations
\be\ba \label{reluvrt}
r&=\sqrt{(r_+^2-r_-^2)(y^2-v^2)+r_+^2},\qquad \phi=\frac{r_+\log \left(\frac{u+x}{u-x}\right)+r_-\log\left(\frac{y+v}{y-v}\right)}{2(r_+^2-r_-^2)}, \\
t& =
\frac{r_-\log \left(\frac{u+x}{u-x}\right)+r_+\log\left(\frac{y+v}{y-v}\right)}{2(r_+^2-r_-^2)}.
\ea\ee
Thus the action of the string propagating in BTZ times $S^1$ is given by
\begin{equation} \label{sigmod}
S = -\frac{\hat \lambda}{2} \int d^2 \sigma
\left( \frac{1}{2} {\rm Tr} ( g^{-1} \partial_a g  g^{-1} \partial^a g^{-1} )
+ \partial_a Z \partial^a Z \right),
\end{equation}
together with the identifications given in (\ref{ident1}).
Here $Z$ is the coordinate along the $S^1$ and $\lambda$ is the coupling of the sigma
model.
Translational symmetry along the time direction $t$ and the angular direction $\phi$ give rise
to the global charges $E$ and $S$. These charges have the following simple
relation in terms of the right and left currents of the sigma model
\be\ba\label{defes}
E + S &= \frac{\hat\lambda}{2} ( r_+- r_-) \int_0^{2\pi} d\sigma {\rm Tr} ( \partial_0 g g^{-1} \sigma^3),\\
E- S &= -\frac{\hat \lambda}{2} ( r_+ + r_-) \int_0^{2\pi} d\sigma {\rm Tr}( g^{-1} \partial_0 g \sigma^3) .
\ea\ee
The equations of motion of the sigma model are given by
\begin{equation}
 \partial^a ( \partial_a g g^{-1}  ) = 0, \qquad  \partial^a\partial_a Z = 0.
\end{equation}
where $a \in \{ 0, 1\}$ refers to the worldsheet coordinates $\tau, \sigma$.
It is convenient to define the light-cone coordinates as
\begin{equation}
 \sigma_{\pm} = \frac{1}{2} ( \tau \pm \sigma) , \qquad \partial_{\pm} = \partial_\tau \pm \partial_\sigma.
\end{equation}
Let us choose a gauge in which
\begin{equation}
 Z = \frac{\hat J}{2\pi \lambda}  \tau + \hat m \sigma,
\end{equation}
where $\hat J$ refers to the momentum on $S^1$ and $\hat m$ refers to the winding.
We will restrict our attention to classical solutions with $\hat m =0$.
Then the Virasoro constraints of the sigma model reduce to
\begin{equation}
\frac{1}{2} {\rm Tr} ( j_\pm^2 ) = - \left( \frac{\hat J}{2\pi \hat \lambda}\right)^2,
\end{equation}
where
\begin{equation}
 \hat  j_\pm = \partial_\pm g g^{-1}.
\end{equation}

Our goal now is to find a method to generate new  solutions for this sigma model
from a given  solution.
For this we adopt the following strategy.
Consider  a solution of the $SL(2, R)$ sigma model
which is
 parameterized as given in (\ref{sl2g}).
We use the isomorphism between $SL(2, R)$ and $SU(1, 1)$ and consider it
as solution in a $SU(1, 1)$ sigma model by the parametrization
\begin{equation}\label{su1g}
 \hat g =
 \left(
 \begin{array}{cc}
   u-iv & x+iy \\
  x-iy & u+iv
 \end{array}
 \right)= Q g Q^{-1},
\end{equation}
where
\begin{equation}
Q=\frac{1}{\sqrt{2}}\left(
 \begin{array}{cc}
   1 & i \\
  1 & -i
 \end{array}
 \right).
\end{equation}
From the fact that $g\in SL(2, R)$ it is easy to show that $\hat g$ satisfies the
defining  property of $SU(1, 1)$ which is given by
\be \label{defsu11}
 \hat g^\dagger M \hat g = M, \qquad {\rm det}\,  \hat g =1, \qquad M = \left( \begin{array}{cc}
            1 & 0 \\
0 & -1
           \end{array}
\right).
\ee
It is clear from the relationship between the  $SU(1, 1)$  and  the $SL(2, R)$
group elements given in (\ref{su1g}) that a solution to the
classical equations of motion  and the Virasoro constraints of the BTZ$\times S^1$ sigma
model will be a solution to the classical equations of motion
and the Virasoro constraints of the  $SU(1, 1)\times S^1$ sigma
model.

Now that we have a solution in the $SU(1, 1)\times S^1$ sigma model we can
use the dressing method developed for this sigma model in \cite{Jevicki:2007pk,Jevicki:2007aa}
to generate new classical solutions to this sigma model. We then
transform it back to a solution in the  BTZ$\times S^1$ sigma model.
This will then be a new solution to the  BTZ $\times S^1$ sigma model.
In general the solutions generated by this method will be classical open strings
moving in the BTZ background.

Before we proceed to explicitly apply this strategy we review the
dressing method for the $SU(1, 1)\times S^1$ sigma model.
Consider the equations of motion to this sigma model which are given by
\begin{equation}\label{eomsu}
\partial_+( \partial_- \hat g \hat g ^{-1} )  + \partial_-( \partial_+ \hat g \hat g^{-1} ) =0.
\end{equation}
To generate new solutions from a given solution we consider the following
system of equations
\be
\label{mon}
i\partial_+\Psi = \frac{A \Psi}{1-\lambda}, \qquad i\partial_-\Psi = \frac{B \Psi}{1+\lambda},
\ee
where $\lambda$ is the complex spectral parameter and
$A, B$ are independent of $\lambda$.
Let us suppose we have a solution to the equations of motion given in (\ref{eomsu}).
Then taking
\begin{equation} \label{absol}
A = i \partial_+ \hat g \hat g^{-1}, \qquad B = i \partial_- \hat g \hat g^{-1},
\end{equation}
guarantees that the integrability constraints of the system of equations in
(\ref{mon}) is satisfied. This is because the equation in  (\ref{eomsu}) together
with the  identity
\begin{equation}
\partial_- ( \partial_+ \hat g \hat g^{-1}) - \partial_+( \partial_- \hat g  \hat g^{-1})
- [ \partial_1 \hat g \hat g^- , \partial_+ \hat g \hat g^{-1} ] =0,
\end{equation}
are the integrability constraints of the system of equations in (\ref{mon}).
Thus for this situation the system (\ref{mon}) can be
solved to obtain $\Psi(\lambda)$  with
\begin{equation}
 \Psi(0) = \hat g.
\end{equation}
Now given the consistent  system of equations in (\ref{mon})  which implies  that the
integrability constraints are satisfied, it is easy to see that $\Psi(0)$ is assured to
satisfy the equation of motion (\ref{eomsu}). This is because the equation of motion is
part of the integrability constraint of the system in (\ref{mon}).
Now to ensure that  we obtain a solution in the group $SU(1, 1)$ we impose the
following constraint
\begin{equation} \label{sucon}
 \Psi^{\dagger}(\bar\lambda) M \Psi(\lambda) = M.
\end{equation}

Given that $A, B$ are  constructed as in (\ref{absol}) from a known solution to (\ref{mon}), we
can generate a new solution by considering a $\lambda$-dependent gauge parameter
$\chi(\lambda)$.  Under this transformation we obtain the system of
equations as in (\ref{mon}) but with $\Psi', A', B' $ given by
\be\ba
\label{trans}
 \Psi '&=\chi \Psi ,\\
A'&=\chi A \chi^{-1} +i(1-\lambda)\partial_+\chi \chi^{-1}, \\
B'&=\chi B \chi^{-1} +i(1+\lambda)\partial_-\chi \chi^{-1} .
\ea\ee
If we ensure that $A', B'$ are independent of $\lambda$, then
it is guaranteed that $\hat g'= \Psi'(0)$ is a possible  new solution to the set of equations
in (\ref{eomsu}). The $SU(1, 1)$ constraint in (\ref{sucon}) requires $\chi$ to satisfy
the equation
\begin{equation}
\chi^\dagger( \bar\lambda) M \chi(\lambda) =M.
\end{equation}

To fix the form of $\chi$ so that $A', B'$ are independent of $\lambda$ we proceed
as follows. The form of $\chi$ is taken to be as
\begin{equation}\label{defchi}
\chi(\lambda) = 1 + \frac{\lambda_1 - \bar\lambda_1}{\lambda - \lambda_1} P,
\end{equation}
where $P$ is a projection operator  given by
\begin{equation} \label{defproj}
P = \frac{\Psi(\bar\lambda_1)e  e^\dagger \Psi^\dagger(\bar\lambda_1) M }{
e^\dagger\Psi^\dagger(\bar\lambda_1) M \Psi(\bar\lambda_1)e},
\end{equation}
where $e$ is a constant vector and $\lambda_1$ is an arbitrary complex parameter.
This projection operators  satisfies the conditions
\begin{equation}
P^2 = P, \qquad M P^\dagger M = P.
\end{equation}
Note that the determinant of $\chi (0)$ is given by
\begin{equation}
{\rm det}  \chi ( 0 ) = \frac{\bar\lambda_1}{\lambda_1}.
\end{equation}
Thus to ensure that the  new solutions which we call the dressed solution belongs to
$SU(1, 1)$, $\hat g'$ is given by
\begin{equation}\label{dressol}
\hat g' = \sqrt{\frac{\lambda_1}{\bar\lambda_1}}  \chi(0 ) \Psi (0) =
\sqrt{\frac{\lambda_1}{\bar\lambda_1}} \chi(0 )\hat g(0).
\end{equation}

We will now show that the form of $\chi$ given in (\ref{defchi}) ensures that
$A', B'$ are independent of $\lambda$.  From our analysis in the previous
paragraphs we see that this is required so that the equations of motion
of the sigma model  is satisfied
by $\Psi'(0)$.
Let us first show that $A'$ given in (\ref{trans}) is independent of $\lambda$.
For this we first evaluate
\begin{equation}
\label{chiui}
\chi^{-1}(\lambda)=1+\frac{\bar{\lambda}_1-\lambda_1}{\lambda-\bar \lambda_1}P.
\end{equation}
From the expression of $A'$ given in (\ref{trans}) and that of
$\chi$ and $\chi^{-1}$ in (\ref{defchi}) and (\ref{chiui}) we see that
the elements of $A'$ are holomorphic functions of $\lambda$ except for the
possible poles at $\lambda = \lambda_1$ and $\lambda= \bar\lambda_1$.
We will  now show that the residue at the possible pole at $\lambda = \lambda_1$
vanishes. Taking the limit $\lambda\rightarrow \lambda_1$ we obtain
\begin{equation}\label{res}
A'(\lambda\rightarrow \lambda_1)
= \frac{\lambda_1 -\bar\lambda_1}{\lambda - \lambda_1}\left[
PA (1-P) + i ( 1-\lambda_1) \partial_+ P ( 1-P)  \right].
\end{equation}
Now using the equations of motion in (\ref{mon}) and the definition of $P$ in
(\ref{defproj}), we see that
\begin{equation}\label{delp}
i\partial_+ P = \frac{AP}{ ( 1- \bar\lambda_1)}  - \frac{PA}{ 1- \lambda_1}
 - \frac{\bar\lambda_1 - \lambda_1}{( 1-\lambda_1) ( 1-\bar\lambda_1)}
 P {\rm Tr}(AP).
 \end{equation}
 Here we have used
 the equation
 \begin{equation}
 A^\dagger M = M A,
 \end{equation}
 which can be obtained by differentiating the $SU(1, 1)$ constraint in (\ref{defsu11}).
 To obtain the last term in (\ref{delp}) we have used the identity
 \begin{equation}
 {\rm Tr} ( AP)  =\frac{ e^\dagger \Psi(\bar\lambda)^\dagger M A \psi(\bar\lambda) e}
 {e^\dagger\Psi^\dagger(\bar\lambda_1) M \Psi(\bar\lambda_1)e},
\end{equation}
which can be shown easily  by taking the trace in the following orthonormal  basis
\begin{equation}\label{comp}
v_1 =
\frac{M \Psi(\bar\lambda_1)e }{\sqrt{ e^\dagger\Psi^\dagger(\bar\lambda_1)  \Psi(\bar\lambda_1)e} }
, \qquad v_2 =  v_{\perp},
\end{equation}
where  $v_\perp$ is the orthogonal unit vector perpendicular to $v_1$.
Using this basis one can also show that ${\rm Tr} P  =1$.
Now substituting  $i \partial_+ P$ from (\ref{delp}) into   the expression for
$A'$ in the limit $\lambda\rightarrow \lambda_1$  given in (\ref{res}),
we see the residue at $\lambda_1$ vanishes.
One can perform the same analysis for the limit $\lambda\rightarrow \bar\lambda_1$ and
show that the potential pole at $\bar\lambda_1$  in $A'$ also vanishes.
This implies that  $A'$ is a meromorphic function in the $\lambda$ plane which approaches the
matrix $A$ at $\lambda\rightarrow \infty$.
Thus Liouville's theorem in complex analysis allows us to conclude that
$A'$ is independent of $\lambda$.  A similar analysis for the matrix $B'$ can
be used to show that $B'$ is independent of $\lambda$.
Therefore the dressed solution given in  (\ref{dressol}) is guaranteed to solve the equations of
motion of the $SU(1, 1)$ sigma model.

\vspace{.5cm}
\noindent
{\bf Dressing preserves Virasoro constraints}
\vspace{.5cm}

We now show that the dressed solution also preserves the Virasoro constraints.
The original solution satisfies the Virasoro constraints which are given by
\begin{equation}
\frac{1}{2} {\rm Tr}   ( \partial_{\pm } \hat g \hat g^{-1} )^2   = - \left(  \frac{\hat J}{2\pi \hat\lambda} \right)^2.
 \end{equation}
 Let us examine the first constraint in the above set of equations, a similar
 analysis applies to the second Virasoro constraint.
From the definition of the dressed solution in (\ref{dressol}) we see that the
dressed solution preserves the Virasoro constraints provided
the following equation is obeyed by the dressing matrix $\chi$
\begin{equation} \label{virp}
{\rm Tr}\left(  2i A \chi^{-1}(0) \partial_{+} \chi(0) + \partial_+ \chi^{-1}(0) \partial_+\chi(0)
\right) =0.
\end{equation}
Now substituting the definition of $\chi$  in  (\ref{defchi}) and its inverse in (\ref{chiui})
as well as the equation (\ref{delp}) we obtain
\begin{equation} \label{man1}
{\rm Tr}\left(  2i A \chi^{-1}(0) \partial_{\pm} \chi(0) \right)
=  \frac{2 ( \lambda_1 -\bar\lambda_1)^2}
{\lambda_1\bar\lambda_1 ( 1- \bar\lambda_1)(1-\lambda_1)}
\left( {\rm Tr}( A^2 P) - ( {\rm Tr}( AP) ) ^2  \right).
\end{equation}
Here we have also used the equation
\begin{equation} \label{sqiden}
{\rm Tr} ( AP AP) = ( {\rm Tr }{AP})^2,
\end{equation}
which holds since the projector in (\ref{defproj}) is a rank 1 projector.
It can also be shown explicitly by  evaluating the trace using the complete set of states
given in (\ref{comp}).
Now using the same manipulations one can show that
\begin{equation}\label{man2}
{\rm Tr} ( \partial_+ \chi^{-1}(0) \partial_+\chi(0) ) = -  \frac{2(  \lambda_1- \bar\lambda_1)^2}
{\lambda_1\bar\lambda_1 ( 1- \bar\lambda_1)(1-\lambda_1)}
\left( {\rm Tr}( A^2 P) - ( {\rm Tr}( AP) ) ^2  \right).
\end{equation}
The equations (\ref{man1}) and (\ref{man2}) imply that the equation
(\ref{virp}) is true and thus the  dressed solution
also preserves the Virasoro constraints.

\section{Dressing time like geodesics}

In this section we apply the dressing method to time like geodesics in
the BTZ background and obtain open string solutions.
The components of the $SL(2, R)$ matrix  for geodesics in BTZ
can be written as
\begin{eqnarray}\label{geosol}
 g_0 = \left( \begin{array}{cc}
  a(\tau) \exp( f(\tau) ) &   b(\tau) \exp( g(\tau)) \\ \nonumber
  b(\tau) \exp( -g(\tau)) & a(\tau) \exp( -f(\tau)) \end{array}
  \right),
  \end{eqnarray}
  with the constraint $a(\tau)^2 - b(\tau)^2 =1$. A convenient parametrization
  for these variables is given by
\begin{equation}
a(\tau) = \cosh\gamma (\tau) , \qquad b(\tau) = \sinh\gamma(\tau).
\end{equation}
The Virasoro constraints for the geodesic reduce to
\begin{equation} \label{virtgeo}
\dot \gamma^2 + \frac{c_1^2}{\cosh^2 \gamma} - \frac{c_2^2}{\sinh^2\gamma}
+ \left( \frac{\hat J}{2\pi \hat \lambda}\right)^2 =0,
\end{equation}
where the dot denotes the  derivative with respect to the worldsheet  time coordinate and
 $c_1, c_2$ are the constants of motion given by
\begin{equation}
\dot f \cosh^2\gamma = c_1, \qquad \dot g \sinh^2\gamma = c_2.
\end{equation}
We choose the initial conditions
\begin{equation}\label{init}
a(0) = a_0, \quad  b(0) = b_0,  \quad f(0) = 0, \quad   b(0) =0,  \quad \dot\gamma(0) = 0.
\end{equation}
It is easy to show using the
equations of motion that  for the geodesic solution given in (\ref{geosol}), the current
\begin{equation}
 j = \partial_+  g_0 g^{-1}_0 = \partial_-  g_0 g^{-1}_0
 \end{equation}
is a constant matrix. Therefore the  solution  to the equation of motion  can also be written as
\begin{equation}
g = \exp( j \tau) \left(
\begin{array}{cc}
a_0 & b_0 \\
b_0 & a_0
\end{array}\right).
\end{equation}
Note that this solution clearly satisfies the initial conditions given in (\ref{init}).
Using this solution consider the  set of equations
\begin{equation} \label{moneq}
\partial_+\Psi_s = \frac{j \Psi_s }{1-\lambda}, \qquad
\partial_-\Psi_s = \frac{j   \Psi_s }{1+\lambda}.
\end{equation}
 Since $j$ is a constant,  the equations in (\ref{moneq}) can easily be integrated.
  Their  solution  is given by
\begin{eqnarray} \label{psisols}
\Psi_s
=\exp\left(j \frac{\tau+\sigma \lambda}{1-\lambda^2}\right) \left(
\begin{array}{cc}
a_0 & b_0 \\
b_0 & a_0
\end{array}\right).
\end{eqnarray}
Note that this solution satisfies the initial condition
\begin{equation}
\Psi_s (0) = g.
\end{equation}
Explicitly performing the exponentiation in the solution (\ref{psisols}) we obtain
\begin{equation} \label{epsisol}
\Psi_s(\lambda) =  \left(
\begin{array}{cc}
a_0 \cos\vartheta + \frac{c_1}{a_0 J} \sin \vartheta  \; \; &
b_0 \cos\vartheta + \frac{c_2}{b_0 J} \sin\vartheta \\
b_0 \cos\vartheta - \frac{c_2}{b_0 J} \sin \vartheta \;\;
&
a_0 \cos\vartheta -\frac{c_1}{a_0 J} \sin \vartheta
\end{array} \right),
\end{equation}
where
\begin{equation}\label{varte}
\vartheta = \frac{J (\tau + \sigma \lambda)}{1-\lambda^2} , \qquad J = \frac{\hat J}{2\pi \hat\lambda}.
\end{equation}
To obtain (\ref{epsisol})  we have also used the Virasoro constraint
\begin{equation}\label{virinpt}
\frac{c_1^2}{a_0^2} - \frac{c_2^2}{b_0^2} + J^2 =0.
\end{equation}

Now given a solution to the set of equations in (\ref{moneq}) it is easy to find the solution
to the corresponding $SU(1,1)$ monodromy equations given in (\ref{mon}).
Since the relation between $SU(1,1)$ and $SL(2, R)$ is given by
(\ref{su1g}) we see that  the corresponding solution to (\ref{mon}) is given by
\begin{equation}\label{tmsusl}
\Psi=   Q \Psi_s Q^{-1}.
\end{equation}
Writing this out explicitly we obtain
\begin{equation} \label{tgeopsi}
\Psi(\lambda) = \left( \begin{array}{cc}
a_0 \cos\vartheta - i \frac{c_2}{b_0 J} \sin\vartheta \;\; &
\frac{c_1}{a_0J} \sin\vartheta + i b_0 \cos\vartheta \\
\frac{c_1}{a_0J} \sin\vartheta - i b_0 \cos\vartheta  \;\; &
a_0 \cos\vartheta + i \frac{c_2}{b_0 J} \sin\vartheta
\end{array}
\right).
\end{equation}
Now that we have the explicit form of $\Psi(\lambda)$ corresponding to the time
like geodesics we can construct $\chi(\lambda)$ as defined in (\ref{defchi}) and proceed to
obtain the dressed solution in the $SL(2, R)$ sigma model which is given by
\begin{equation}\label{dress1}
 g' = Q^{-1} \hat g' Q  = Q^{-1} \sqrt{\frac{\lambda_1}{\bar\lambda_1}}\chi(0) \hat g (0)  Q,
\end{equation}
where we have substituted  $\hat g'$  from  (\ref{dressol}).

\subsection{The dressed solution}

In this section we will present an explicit example of a class of classical string
solutions obtained by applying the dressing method on time like geodesics.
For this we will first simplify the situation by considering time like geodesics
which satisfy the following relation
\be
\label{limit}
c_1 = c_2 = a_0b_0J.
\ee
The Virasoro constraint at the initial point $\tau=0$ given in (\ref{virinpt})
 is trivially satisfied with the condition given in (\ref{limit}).
Evaluating the global charges of these time like geodesics we obtain the following
\begin{equation}\label{cggeo}
 E+ S = 0, \qquad  E -S =  - 2\pi \hat \lambda( r_+ + r_-) ( c_1 + c_2) = -4\pi a_0b_0 J.
\end{equation}
Using the condition (\ref{limit}) in
the  expression for  $\Psi(\lambda)$  given in (\ref{tgeopsi}) we find that it reduces to
\begin{equation}
\Psi(\lambda) = \left( \begin{array}{cc}
a_0 \exp( -i \vartheta) & i b_0 \exp ( - i \vartheta) \\
-i b_0 \exp( i \vartheta) & a_0 \exp( i \vartheta)
\end{array}
\right).
\end{equation}
We will now choose the following constant vector for constructing the dressing factor
\begin{equation}
e = \left( \begin{array}{c}
1 \\ 0 \end{array}
\right).
\end{equation}
Using the definition of $P$ given in (\ref{defproj}) we find its  elements
are given by
\be\ba\label{proje}
 P_{11} &= \frac{a_0^2 \exp ( -J_1)}{ a_0^2 \exp( -J_1) - b_0^2 \exp( J_1) },
\quad
&P_{12} &= \frac{ -i a_0b_0 \exp( -iJ_1) }{ a_0^2 \exp( -J_1) - b_0^2 \exp( J_1) },
\\
P_{21} &= \frac{ -i a_0b_0 \exp( iJ_1) }{ a_0^2 \exp( -J_1) - b_0^2 \exp( J_1) },
\quad
&P_{22} &=  \frac{-b_0^2 \exp ( J_1)}{ a_0^2 \exp( -J_1) - b_0^2 \exp( J_1) },
\ea\ee
where
\be\ba
\label{j1}
iJ_1&=J\left[\frac{\tau+\sigma \lambda_1}{1-\lambda_1^2}-\frac{\tau+\sigma \bar\lambda_1}{1-\bar\lambda_1^2}\right]
=i2J\left[\frac{\tau r^2 \sin(2\theta)+r\sigma\sin(\theta)(1+r^2)}{1+r^4-2r^2\cos(2\theta)}\right],  \\
J_2&=J\left[\frac{\tau+\sigma \lambda_1}{1-\lambda_1^2}+\frac{\tau+\sigma \bar\lambda_1}{1-\bar\lambda_1^2}\right]
=2J\left[\frac{\tau(1-r^2\cos(2\theta))+\sigma r \cos(\theta)(1-r^2)}{1+r^4-2r^2\cos(2\theta)}\right].
\ea\ee
Note that the projection matrix has the following property:
Given a definite value of $\tau$, the asymptotic values of $P$ for $\sigma \rightarrow \pm\infty$
are given by
\begin{equation}\label{projas}
 P(\tau, \infty) = \left(\begin{array}{cc}
                          0 & 0 \\
0 & 1
                         \end{array}\right),
\quad
P(\tau, -\infty) = \left(\begin{array}{cc}
                          1 & 0 \\ 0 & 0
                         \end{array}
\right).
\end{equation}
By using the projection matrix given in
(\ref{proje}) to construct  the normalized dressing factor
dressing factor $\hat \chi(0 )$  we obtain
\be\ba
\label{chisu}
 \hat \chi_{11}(0)=
\exp(i\theta)-\frac{2i\sin(\theta)a_0^2\exp(-J_1)}{a_0^2\exp(-J_1)-b_0^2\exp(J_1)}, &\quad&
\hat \chi_{12}(0)=-\frac{2\sin(\theta)a_0b_0\exp(-iJ_2)}{a_0^2\exp(-J_1)-b_0^2\exp(J_1)}, \\
\hat \chi_{22}(0)=\exp(i\theta)+\frac{2i\sin(\theta)a_0^2\exp(-J_1)}{a_0^2\exp(-J_1)-b_0^2\exp(J_1)},
&\quad&
\hat \chi_{21}(0)=-\frac{2\sin(\theta)a_0b_0\exp(iJ_2)}{a_0^2\exp(-J_1)-b_0^2\exp(J_1)} ,
\ea\ee
where
\begin{eqnarray}
\hat \chi &=& \sqrt{\frac{ \lambda_1}{\bar \lambda_1}} \chi(0),
\end{eqnarray}
and the complex parameter $\lambda_1 = r e^{i \theta}$. It can be easily verified that the
dressing factor in (\ref{chisu}) belongs to $SU(1, 1)$.  Note that when the parameter $\lambda_1$ becomes
real, that is $\theta =0$, the dressing factor reduces to identity. Thus there is a smooth
limit in which the dressed solution will reduce to the original geodesic.
Now using  the  expression in (\ref{dress1}), the new solution to the
sigma model which satisfies the Virasoro constraints is given by
\be\ba
\label{dressedsol}
u'&=a_0\cos(\theta-J\tau)-\frac{2\sin(\theta)(a_0^3\exp(-J_1)\sin(J\tau)+a_0b_0^2\sin(J\tau-J_2))}{a_0^2\exp(-J_1)-b_0^2\exp(J_1)},\\
v'&=-a_0\sin(\theta-J\tau)+\frac{2\sin(\theta)(a_0^3\exp(-J_1)\cos(J\tau)-a_0b_0^2\cos(J\tau-J_2))}{a_0^2\exp(-J_1)-b_0^2\exp(J_1)},\\
x'&=-b_0\sin(\theta-J\tau)+\frac{2\sin(\theta)a_0^2b_0(\exp(-J_1)\cos(J\tau)-\cos(J\tau-J_2))}{a_0^2\exp(-J_1)-b_0^2\exp(J_1)},\\
y'&=b_0\cos(\theta-J\tau)-\frac{2\sin(\theta)a_0^2b_0 (\exp(-J_1)\sin(J\tau)+\sin(J\tau-J_2))}{a_0^2\exp(-J_1)-b_0^2\exp(J_1)} ,
\ea\ee
where $J_1$ and $J_2$ are given by (\ref{j1}).
We have used the parametrization of $SL(2, R)$ given in (\ref{sl2g}) to read out
the  $u', v', x', y'$  values of the dressed solution.
Note that as a simple check it can be verified that the constraint
$-u^{\prime 2} - v^{\prime 2} + x^{\prime 2} + y^{\prime 2} =-1$
is satisfied. We have also explicitly verified that the solution
in (\ref{dressedsol}) satisfies both the equation of motion as well
as the Virasoro constraints given by
\begin{equation}
 \label{virtime}
 -\partial_{\pm}(u'+x')\partial_{\pm}(u'-x')+
\partial_{\pm}(y'+v')\partial_{\pm}(y'-v')= -J^2.
\end{equation}
Multiple dressed solutions can be obtained by the general procedure discussed in
the appendix.

\vspace{.5cm}\noindent
{\bf Description of the solution}
\vspace{.5cm}

We now briefly describe the features of the solution. At any given value of the worldsheet time
the end points of the string $\sigma \rightarrow \pm \infty$
move as time like geodesics. Taking  the limit $\sigma\rightarrow \infty$ in
the solution given in (\ref{dressedsol}), we obtain
\be\ba \label{geo1}
u'  &= a_0 \cos( \theta - J\tau) , &\qquad& v' = -a_0 \sin (\theta - J\tau), \\
x' &= -b_0 \sin(\theta - J\tau) , &\qquad& y'= b_0 \cos(\theta - J\tau).
\ea\ee
In taking this limit we have assumed that $\sin\theta >0$ (a similar analysis can
be repeated for $\sin\theta<0$).
Comparing this to the seed solution $\Psi_x(0) = g$ given in (\ref{epsisol}) we see that it is the
 geodesic with same constants of motion  but with different initial conditions.
Similarly taking the limit $\sigma\rightarrow -\infty$ we obtain
\be\ba
u' &= a_0 \cos(\theta + J \tau) , &\qquad& v' = a_0 \sin(\theta + J \tau), \\
x' &= b_0 \sin(\theta + J\tau), &\qquad& y' = b_0 \cos(\theta + J\tau).
\ea\ee
Again, this is the same geodesic as in (\ref{geo1})  with $\theta \rightarrow -\theta$.
Thus, the end points of the string move as geodesics with the same
constants of motion as the seed geodesic, but with initial conditions depending on
$\theta$.

To further characterize these solutions we will evaluate their global charges.
We first show that the charge $E+S =0$, thus preserving the condition of the seed geodesic
given in (\ref{cggeo}). The charge of the dressed solution is given by
\be\ba \label{lcha}
 \frac{E + S}{r_+ - r_-} &= \frac{\hat \lambda}{2} \int_{-\infty}^\infty d\sigma
{\rm Tr} (\partial_0 g' g^{\prime -1}  \sigma^3 ) , \\
&=  \frac{\hat \lambda}{2} \int_{-\infty}^\infty d\sigma
{\rm Tr }(\partial_0\hat g^{\prime} \hat g^{\prime -1} Q\sigma^3 Q^{-1} ), \\
&=  \frac{\hat \lambda}{2} \int_{-\infty}^\infty d\sigma
{\rm Tr } (\partial_0\hat g^{\prime} \hat g^{\prime -1} \sigma^1 ).
\ea\ee
In the second and third line of the equation we have converted the $SL(2, R)$ variables
to $SU(1, 1)$. Now from the fact that the dressing method is a gauge transformation
the new currents are related to the old ones by (\ref{trans})
\be\ba
A'&=\chi A \chi^{-1} +i(1-\lambda)\partial_+\chi \chi^{-1}, \\
B'&=\chi B \chi^{-1} +i(1+\lambda)\partial_-\chi \chi^{-1}.
\ea\ee
We have also seen that $A', B'$ are independent of $\lambda$, therefore we
can evaluate the LHS of the above equations at any convenient value of
$\lambda$. Let us evaluate the LHS at $\lambda\rightarrow\infty$.
This results in the following equations
\be\ba
 \partial_+ \hat g' \hat g^{\prime -1} &= \partial_+ \hat g \hat g ^{-1}
 - ( \lambda_1 -\bar\lambda_1)  \partial_+ P, \\
\partial_- \hat g'\hat g ^{\prime -1} &= \partial_- \hat g \hat g^{-1}
+  (\lambda_1 - \bar\lambda_1) \partial_- P.
\ea\ee
Adding both equations we obtain the following
\begin{equation}
\partial_0 \hat g' \hat g^{\prime -1} = \partial_0 \hat g \hat g^{-1}
 - ( \lambda_1 - \bar \lambda_1) \partial_\sigma P.
\end{equation}
 This equation is very convenient to evaluate the charge given in
(\ref{lcha}) .
\be\ba \label{rcharge}
\frac{2( E+ S) }{\hat \lambda( r_+ -r_-) }   &=  \int_{-\infty}^\infty d\sigma
{\rm Tr } \left( \partial_0\hat g \hat g^{ -1} \sigma^1 \right)
 - (\lambda_1 - \bar\lambda_1) {\rm Tr}
\left[( P(\tau, \infty) - P(\tau, -\infty) \sigma^1 \right]   \\
&= 0.
\ea\ee
To obtain the last line we have used the fact that the seed geodesic has the
property $c_1 = c_2$ and the values of the asymptotic values of the projection
matrix are given in \eqref{projas}.
Thus the dressing does not change the left charge $E+S$.
It is easy to see from the analysis that this holds true for any number of
dressings of this geodesics provided the asymptotic property of the
projection matrix is that given in \eqref{projas}.
Now let us study the charge $E-S$.  Here we need to explicitly perform the
integrals. By a tedious calculation,
it can be shown that  the following integral is given by
\be\ba
\int Tr(g^{\prime -1}\partial_\tau g^{\prime}   \sigma^3)d\sigma&=4a_0b_0\left[J\sigma+
\frac{-2a_0^2+(1+2b_0^2)\exp(2 B)\cos( C)}{(-1+b_0^2(-1+\exp(4 B)))r}
\sin(\theta)\right], \\
B&=\frac{Jr[\sigma(1+r^2)+2\tau r \cos(\theta)]}{1+r^4-2r^2\cos(2\theta)}\sin\theta ,\\
C&=\frac{2Jr[\sigma(r^2-1)\cos(\theta)+r\tau(r^2-\cos(2\theta))]}{1+r^4-2r^2\cos(2\theta)}.
\ea\ee
Now on taking the limits
we obtain the following global charge
\be\ba
\label{lch}
E-S&= -\frac{\hat \lambda}{2}(r_++r_-)\lim_{L\rightarrow\infty} \int_{-L}^{L}d\sigma Tr(g^{-1}\partial_\tau g  \sigma^3) , \\
&= -\frac{\hat \lambda}{2}(r_++r_-)(8a_0b_0JL-8\frac{a_0b_0|\sin \theta|}{r}).
\ea\ee
Thus the leading contribution to the global charge diverges linearly with
the worldsheet length.
From (\ref{cggeo}) we see that the uniform charge density which contributes to this
divergence is the same as that of the seed geodesic.
Therefore we regulate the charges by simply subtracting this charge density.
Since $E=-S$ from (\ref{rcharge}) we see that we must define the following
regulated charges
\begin{equation}
 \hat E = E + 2 \hat \lambda(r_++ r_-) a_0b_0 JL ,  \qquad
\hat S = S - 2  \hat \lambda(r_++ r_-) a_0b_0 JL.
\end{equation}
In terms of these regulated charges we obtain the following dispersion relation
\begin{equation}
 \hat E - \hat S = \hat \lambda(r_++r_-) \frac{4a_0 b_0 |\sin\theta|}{r}.
\end{equation}
The dispersion relation of these classical solutions resembles that of the
giant magnon.

For the case $r=0$, the solution reduces to a geodesic but with different
conserved charges.  Substituting $r=0$ in (\ref{j1}) and using the result in
the equations (\ref{dressedsol}) we obtain the following geodesic
\be\ba
u' &= a_0 \cos (\theta + J\tau) , &\qquad&  v' = a_0\sin(\theta + J\tau), \\
x' &= - b_0\sin(\theta - J\tau) , &\qquad& y' = \cos(\theta - J\tau).
\ea\ee
Evaluating the global charges for this case we obtain
\begin{equation}
E+ S = 0, \qquad E-S = - 4\pi \hat\lambda( r_+ + r_-) a_0b_0 \cos(2\theta).
\end{equation}

\begin{figure}
\centering
\includegraphics[width=70mm,height=50mm]{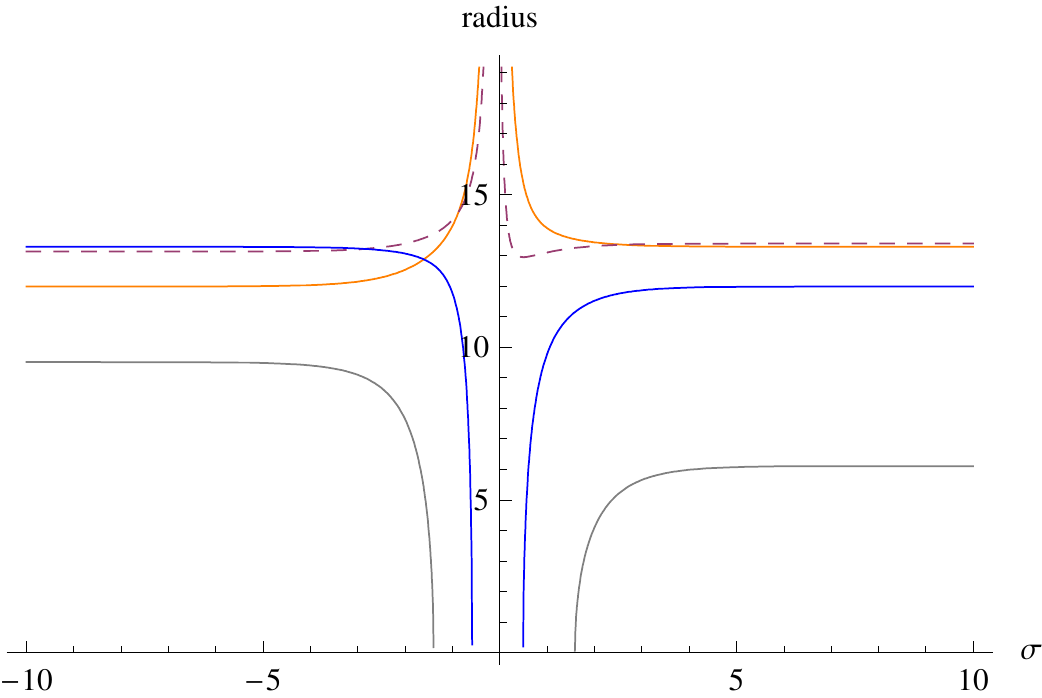}
\includegraphics[width=70mm,height=50mm]{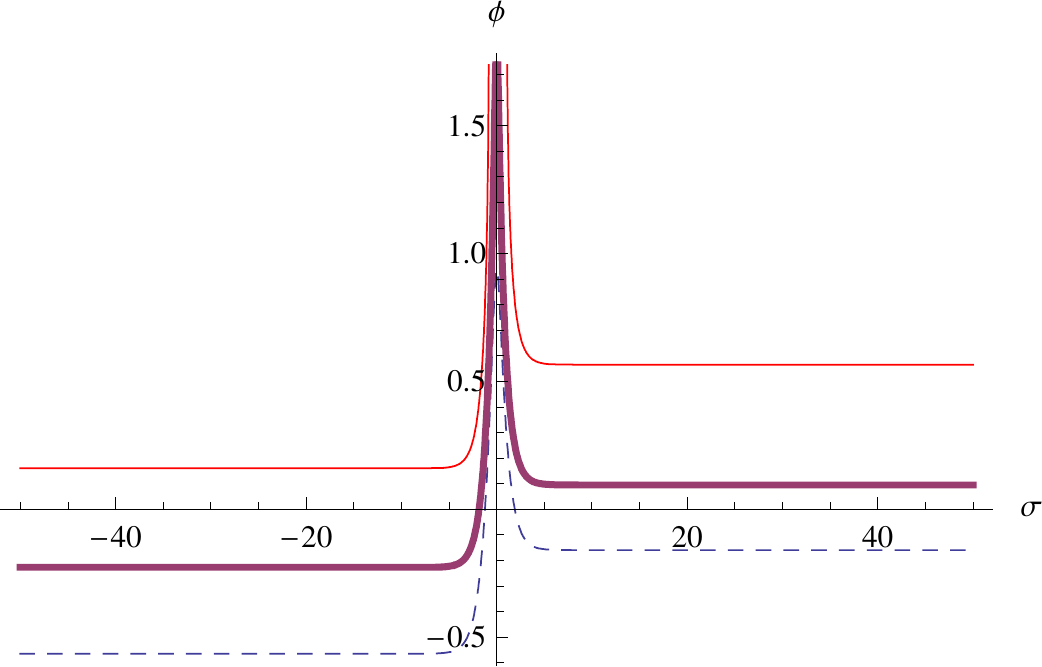}
\includegraphics[width=70mm,height=50mm]{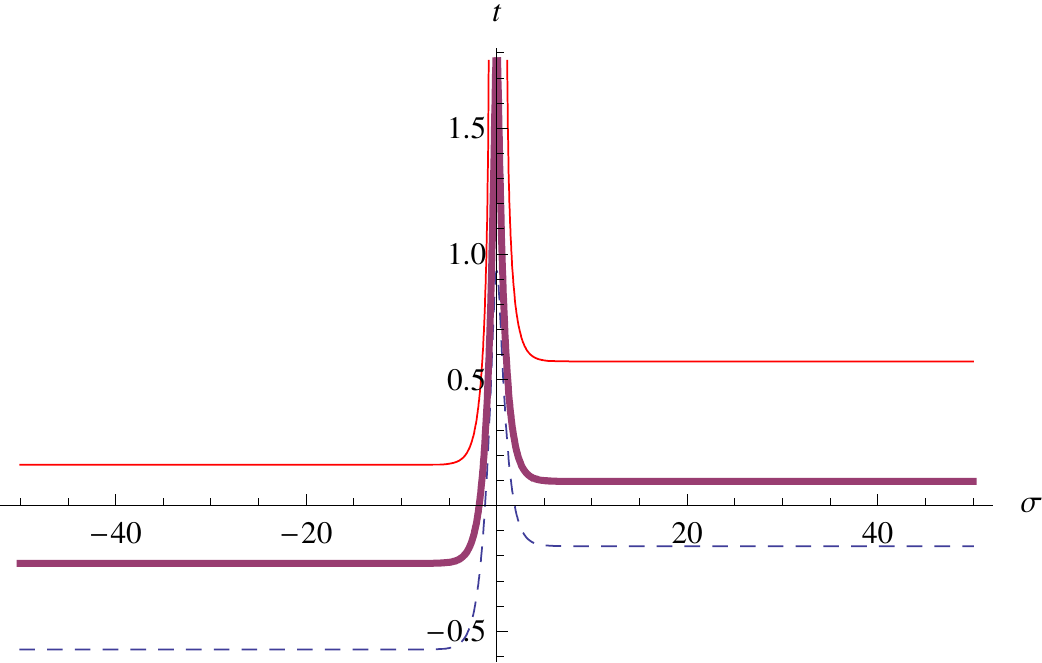}
\caption{Orange ($\tau=1.0$), dashed ($\tau=1.2$), thick ($\tau=1.5$), grey ($\tau=2.0$).}
\label{fig1}
\end{figure}

\begin{figure}
\centering
\includegraphics[width=100mm,height=70mm]{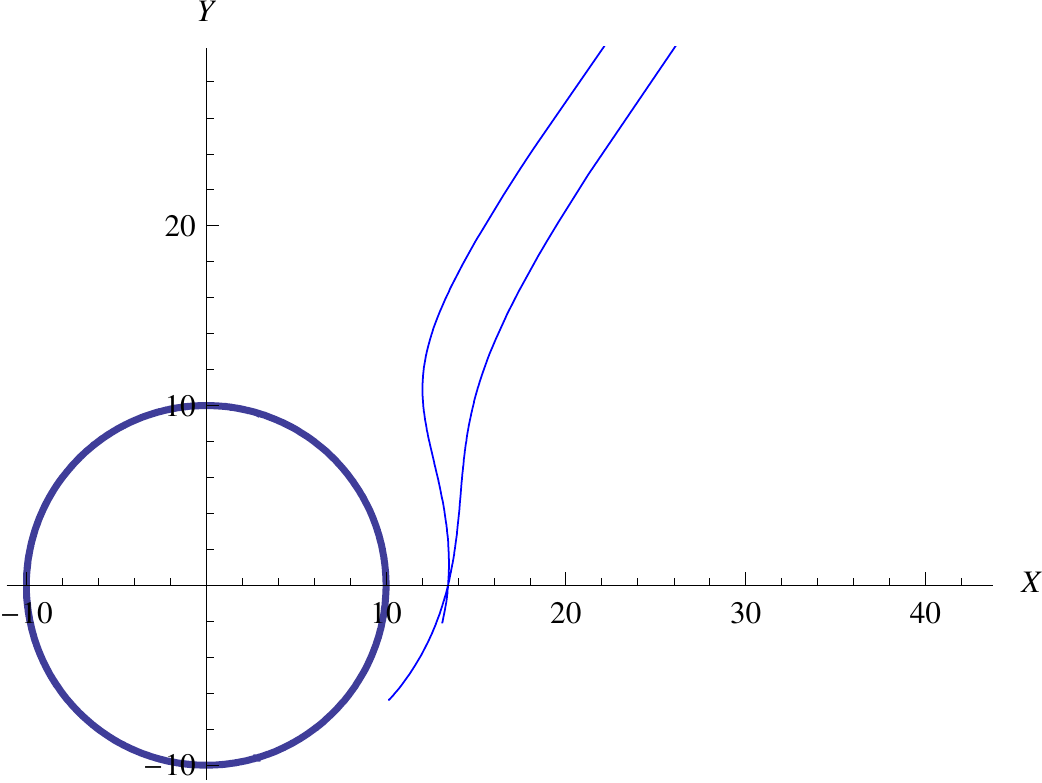}
\caption{String in $r-\phi$ plane.}
\label{fig2}
\end{figure}

\begin{figure}
\centering
\includegraphics[width=100mm,height=70mm]{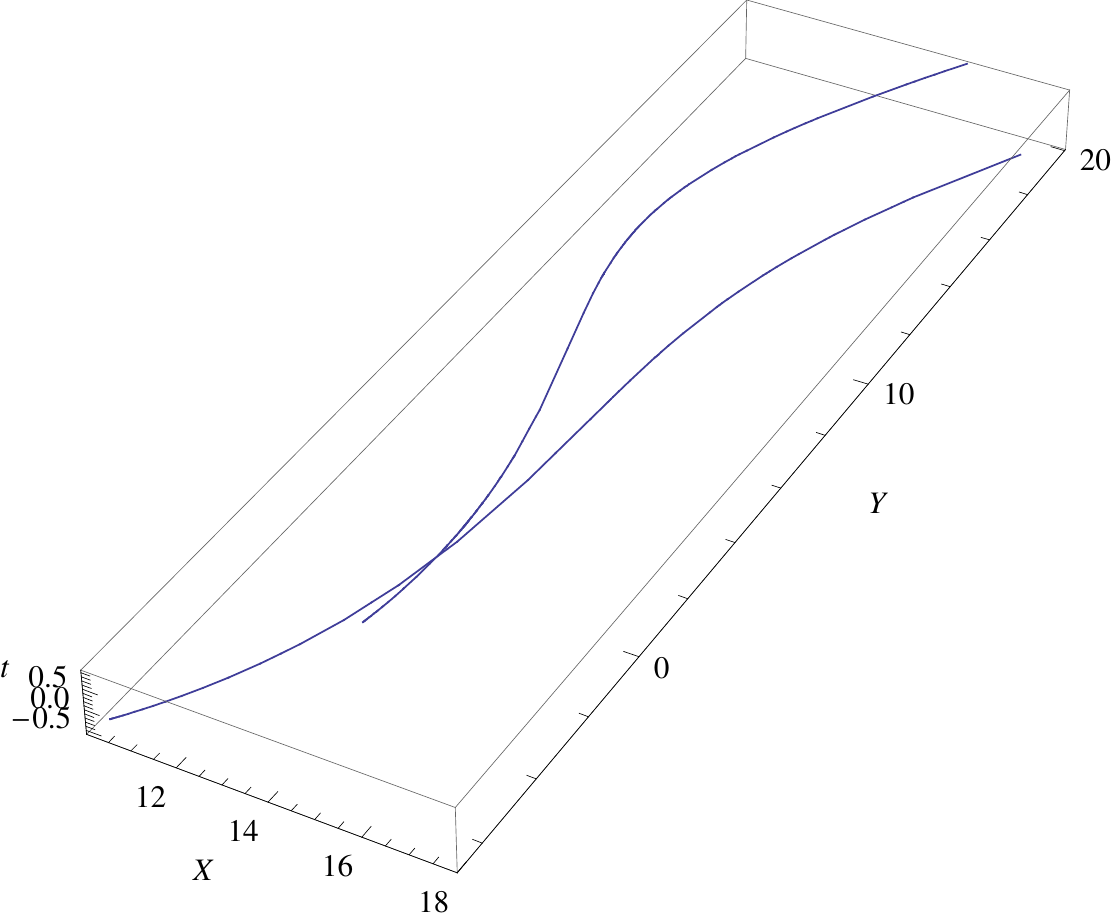}
\caption{String in 3d BTZ background.}
\label{fig3d}
\end{figure}

We now describe the snapshot of the string at a given value of the worldsheet time
$\tau$.
From the expression for the coordinates  given in (\ref{dressedsol})
 we see that in general they acquire
large values when the following equality is satisfied
\begin{equation}\label{spike}
e^{-J_1} = \frac{b_0}{a_0}.
\end{equation}
Using the  expression for $J_1$ in (\ref{j1}) we see that given a particular value of
$\tau$, the equality in (\ref{spike})
will be satisfied at one point say $\sigma^*$.
From the relations between the embedding coordinates $u, v, x, y$ and the
BTZ coordinates $r, \theta, t $ given in (\ref{reluvrt}) we see that  at
$\sigma^*$  the string possibly  is
at the boundary of the BTZ geometry.  As discussed before the end points of the string move
on time like geodesics. Thus at a given value of $\tau$ is pinned at the boundary and the
end points lie on time like geodesics.
An estimate of the worldsheet time at which the string falls into the horizon is given by
\begin{equation}
\tau_{\rm fall}  = {\rm Min}
\left( \frac{1}{J}\left | \tan^{-1} \frac{b_0}{a_0} + \theta \right|,
\frac{1}{J}\left | \tan^{-1} \frac{b_0}{a_0} - \theta\right |  \right),
\end{equation}
where the ${\rm Min}$ refers to the minimum  of  the quantities in the
bracket.
This estimate is obtained from the time at which the end points reach the horizon.
We now plot the snapshot of the solution for some typical
values of the parameters in the BTZ geometry. In figure \ref{fig1}  we have plotted
the radial position of the solution
(\ref{dressedsol})  against the worldsheet $\sigma$ coordinate for the following
set of parameters.
\be
\label{par}
b_0=\frac{0.9}{\sqrt{0.19}},\quad r_+=10,\quad r_-=9,\quad J=\frac{0.8 \pi}{2},\quad \theta=\frac{0.9 \pi}{2}, \quad r=1.
\ee
The above choice of $b_0$ gives $b_0/a_0=0.9$.
 Note that $r$ and $\phi$ reach asymptotic values at large values at $\sigma^*$
as predicted by the solution.
The string is completely outside the horizon which is at $r_+=10$ for worldsheet time $\tau$=1.
As $\tau$  increases we see that the end points cross the horizon and the point which
is pinned at the boundary falls in.
In figure \ref{fig2}   we have plotted the projection of the snapshot in the
$r-\phi$ plane for $\tau=1$, the range of $\sigma$ chosen for the plot is from $-5$ to $5$.
The two end points near the horizon are the ends of the strings which
move on time like geodesics.
The strings extends along the almost parallel lines which meet at the boundary
of BTZ.  Figure \ref{fig3d}  shows the  string  configuration
at $\tau=1$ in the 3 dimensional BTZ spacetime.

\section{Dressing space like geodesics}

Space like geodesics in the BTZ background play an important role in the framework of
$AdS_3 /CFT_2$ . The regularized length of the space like geodesic connecting two equal time
points on the boundary is a useful WKB estimate of the correlator of operators of large
dimension operators inserted at these points  \cite{Louko:2000tp,Balasubramanian:1999zv,Kraus:2002iv,Fidkowski:2003nf,Festuccia:2005pi}.
Their lengths are also used to determine
Wilson loops and entanglement entropy in the boundary theory
\cite{Nishioka:2009un,Hubeny:2012ry}.
In the $SL(2, R)$ WZW
model they played an important role in generating new solutions using spectral flow
 \cite{Maldacena:2000hw}.
In this section we study the solutions to the sigma model in the BTZ background obtained by
dressing space like geodesics. Space like geodesics which ends on two points of the boundary
never penetrate the horizon of the black hole. Starting at a point in the boundary they
reach a minimum distance and before the turn back to the boundary. An interesting result
we obtain is that the minimal surfaces obtained by dressing these space like geodesics can
penetrate the horizon while being pinned on some curve at the boundary. Thus these can
possibly be good probes of the physics at the horizon.

\begin{figure}
\centering
\includegraphics[width=80mm,height=50mm]{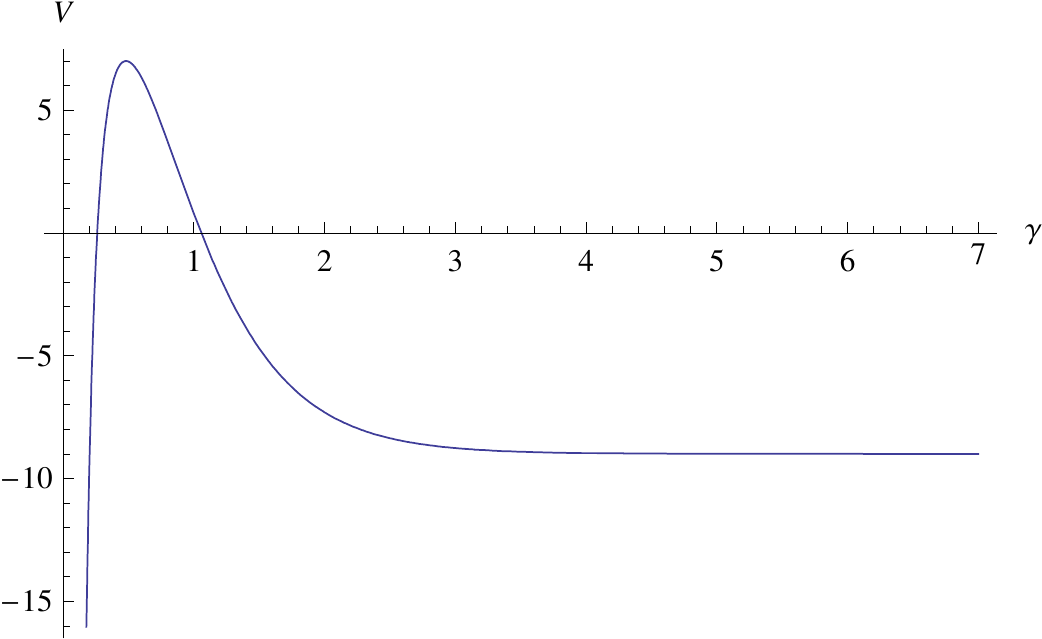}
\caption{V($\gamma$) vs $\gamma$ plot for $c_1=5$, $c_2=1$, $J=3$.}
\label{hump}
\end{figure}

\begin{figure}
\centering
\includegraphics[width=80mm,height=50mm]{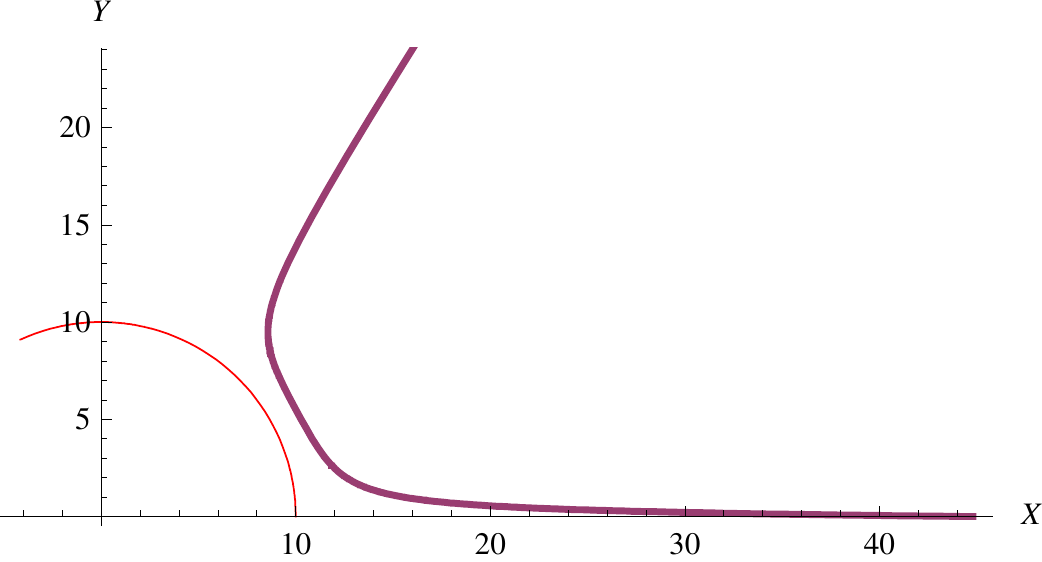}
\caption{Trajectory of the tachyon for $c_1=5$, $c_2=1$, $J=3,~r_+=10,~r_-=9$ in $r-\phi$ plane of BTZ.}
\label{figeo}
\end{figure}

For the time like geodesics  the mass of the particle was identified
as the momentum  $J$ of the circle
in the $BTZ \times S^1$ sigma model given in (\ref{sigmod}).
For the space like geodesics the equation of motion remains the same however
the Virasoro constraints are changed.  They are given by
\be\label{slvir}
\dot \gamma^2 + \frac{c_1^2}{\cosh^2 \gamma} - \frac{c_2^2}{\sinh^2\gamma}
- J^2 = 0.
\ee
Comparing this constraint with that corresponding to  the time like case
given in (\ref{virtgeo}) we see that for the space like geodesics  we need
to replace  $J^2  \rightarrow -J^2$.
We are now interested in geodesics that can originate from the boundary
and reach back to the boundary.
For this let us examine the potential for the space like geodesics. It is given by
\be
V(\gamma)=\frac{c_1^2}{a^2}-\frac{c_2^2}{b^2}-J^2.
\ee
For definiteness let us assume $c_1, c_2, J \geq 0$. Then the maximum of the
potential occurs at
\begin{equation}
\sinh \gamma  = \sqrt {\frac{ c_2}{ c_1 - c_2} }.
\end{equation}
This maximum exists for $c_1>c_2$. For the geodesic to turn back to the boundary
we need the value at this  maximum to be greater than zero.
This is achieved by the following condition
\begin{equation}
 c_1 - c_2 > J.
\end{equation}
Once these  conditions are satisfied it is clear  that
a geodesic starting at the boundary will return back to the boundary.
It will never reach the horizon which is at $\gamma =0$.
An example of values which satisfy these conditions
are $c_1 =5,~ c_2 = 1,~ J=3$. The potential is plotted in figure \ref{hump}.
A trajectory of a space like geodesic originating close to the boundary and going back to the boundary
after hitting the hump located at
\begin{equation}
 \gamma_c=\sinh^{-1}\left( \sqrt{\frac{c_2}{c_2-c_1}}\right) =0.4812
\end{equation}
is plotted in the $r-\phi$ plane in figure \ref{figeo}.

It is clear from the generic shape of the potential for  the condition $c_1-c_2>J$, that the
space like geodesic never reaches the horizon. We will show that upon dressing these space
like geodesics  we will obtain classical string solutions  that can penetrate the horizon while
being pinned on the boundary at two points. The reason this is possible is perhaps due
to the fact that   the worldsheet in this case has Minkowski signature unlike the situation discussed
in \cite{Hubeny:2012ry} where is was shown that Euclidean minimal surfaces pinned
at the boundary do not penetrate horizons.
We will also  show that there is a domain of parameter space in which
these solutions are closed strings.
 Just as  space like geodesics which originate and
end at the boundary can be used as probes of horizon physics it is plausible that
these solutions can give information of the physics of the horizon.

To solve for the trajectory of a general space like geodesic we again use the ansatz
given in (\ref{geosol}).  The initial conditions are given by
\begin{equation} \label{inispace}
 a(0) = a_0, \quad b(0) = b_0, \quad  f(0) = 0, \quad b(0) = 0, \quad
\dot \gamma(0)  = - \sqrt{J^2+\frac{c_2^2}{b_0^2}-\frac{c_1^2}{a_0^2}}.
\end{equation}
We start at some large $\gamma_0$ outside the maxima of the potential with
radially inward velocity.
The solution to the monodromy equations given in (\ref{moneq})
can be obtained by following the similar procedure adopted for
time like geodesics. The results are
\begin{eqnarray}
\Psi_s = \left( \begin{array}{cc}
\Psi_{s( 11)} & \Psi_{s(12)}\\
\Psi_{s(21)} & \Psi_{s(22)}
\end{array}
\right),
\end{eqnarray}
where
\be\ba
\Psi_{s(11)} &= a_0\cosh[\vartheta]-\frac{b_0\sqrt{J^2+\frac{c_2^2}{b_0^2}-\frac{c_1^2}{a_0^2}}}{J}
\sinh[\vartheta]
+ \frac{c_1}{a_0J}\sinh[\vartheta],   \\
\Psi_{s(12)} &= b_0\cosh[\vartheta]-\frac{a_0\sqrt{J^2+\frac{c_2^2}{b_0^2}-\frac{c_1^2}{a_0^2}}}{J}
\sinh[\vartheta] + \frac{c_2}{b_0J}\sinh[\vartheta] , \\
\Psi_{s(21)} &=   b_0\cosh[\vartheta]-\frac{a_0\sqrt{J^2+\frac{c_2^2}{b_0^2}-\frac{c_1^2}{a_0^2}}}{J}
\sinh[\vartheta] - \frac{c_2}{b_0J}\sinh[\vartheta] , \\
\Psi_{s(22)} &=  a_0\cosh[\vartheta]-\frac{b_0\sqrt{J^2+\frac{c_2^2}{b_0^2}-\frac{c_1^2}{a_0^2}}}{J}
\sinh[\vartheta]
- \frac{c_1}{a_0J}\sinh[\vartheta],
\ea\ee
and $\vartheta$  and $J$ are  defined in (\ref{varte}).
Note that this solution satisfies the condition
\begin{equation}
\Psi_s(0) = g,
\end{equation}
where $g$  refers to the trajectory of the space like geodesic with the initial conditions
given in (\ref{inispace}).  To obtain this form of the monodromy matrix we have
used the Virasoro constraint given in (\ref{slvir}).  Using
the transformation given in  (\ref{tmsusl})   the solution to the
$SU(1, 1)$ monodromy equations can be constructed.
To write this solution in a convenient form we define
\be
c_1=n_1a_0J, \qquad c_2=n_2b_0J,
\ee
where $n_1$ and $n_2$ are arbitrary parameters. The
solution to  the $SU(1,1)$ monodromy equations in   (\ref{mon}) is then given by
\be\ba
\Psi_{11}&= \cosh\gamma_0\cosh\vartheta -c\sinh(\gamma_0)\sinh \vartheta -
in_2\sinh \vartheta, \\
\Psi_{12}&= n_1\sinh\vartheta+i\left(\sinh\gamma_0\cosh\vartheta-
c\cosh\gamma_0\sinh\vartheta \right), \\
\Psi_{21}&=n_1\sinh\vartheta -i\left(\sinh\gamma_0
\cosh\vartheta -c\cosh\gamma_0\sinh\vartheta \right), \\
\Psi_{22} &= \cosh\gamma_0\cosh\vartheta -c\sinh(\gamma_0)
\sinh\vartheta +in_2\sinh\vartheta , \\
\ea\ee
where
\begin{equation}
 c=\sqrt{1-n_1^2+n_2^2}.
\end{equation}

We now  proceed as in the case of the time
like geodesics and construct the dressing factor
given in (\ref{defchi}) with the constant vector $e=\left(
 \begin{array}{cc}
   1\\
  0
 \end{array}
 \right)$.
The dressed solution in the $SL(2, R)$ sigma model is then obtained by
using (\ref{dress1}). This results in the following solution
\be\ba
\label{u1s}
u'=&u\cos(\theta)+ \\
& \sin(\theta)\left[-\sinh(\gamma_0)(e^{\gamma '} \sinh(J\tau-2\alpha)+e^{\tilde \gamma}c\cosh(J\tau-2\alpha))+ \right. \\
& \left. (c\sinh(\gamma_0)\cosh(J\tau)-\sinh(J\tau)\cosh(\gamma_0))(e^{\tilde \gamma}\cos(2\beta)+\sin(2\beta))\right]\\
&\times(\cos(2\beta)-\exp(\tilde \gamma)\sin(2\beta))^{-1}, \\
v'=&v\cos(\theta)  \\
& + \sin(\theta)\left[ \cosh(J\tau-2\alpha)(\cosh(\gamma_0)+n_2e^{\tilde \gamma})+c\sinh(\gamma_0)\sinh(J\tau-2\alpha) \right. \\
& \left. -n_2\cosh(J\tau)(e^{\tilde \gamma}\cos(2\beta)+\sin(2\beta))\right](\cos(2\beta)-\exp(\tilde \gamma)\sin(2\beta))^{-1} , \\
x'=&x\cos(\theta)   \\
& + \sin(\theta)\left[ \cosh(J\tau-2\alpha)(n_1e^{\tilde \gamma}-\sinh(\gamma_0))-c\cosh(\gamma_0)\sinh(J\tau-2\alpha) \right. \\
& \left. -n_1\cosh(J\tau)(e^{\tilde \gamma}\cos(2\beta)+\sin(2\beta))\right](\cos(2\beta)-\exp(\tilde \gamma)\sin(2\beta))^{-1} ,\\
y'=&y\cos(\theta)+ \nonumber  \\
& \sin(\theta)\left[-\cosh(\gamma_0)(e^{\gamma '} \sinh(J\tau-2\alpha)+e^{\tilde \gamma}c\cosh(J\tau-2\alpha))+ \right. \\
& \left. (c\cosh(\gamma_0)\cosh(J\tau)-\sinh(J\tau)\sinh(\gamma_0))(e^{\tilde \gamma}\cos(2\beta)+\sin(2\beta))\right]\\
&\times(\cos(2\beta)-\exp(\tilde \gamma)\sin(2\beta))^{-1}, \\
\ea\ee
where
\begin{equation}
e^{\tilde \gamma} =n_2\cosh(\gamma_0)+n_1\sinh(\gamma_0), \qquad e^{\gamma'}=n_1\cosh(\gamma_0)+n_2\sinh(\gamma_0),
\end{equation}
and
\begin{equation}\label{alp}
 \alpha = \frac{\tau(1-r^2\cos(2\theta))+
\sigma \cos(\theta)(1-r^2)}{1+r^4-2r^2\cos(2\theta)}J, \quad
 \beta= \frac{\tau r^2 \sin(2\theta)
+r\sigma\sin(\theta)(1+r^2)}{1+r^4-2r^2\cos(2\theta)}J.
\end{equation}
Note that $\alpha = J_2/2$ and $\beta = J_1/2$ as defined in (\ref{j1}).
We have also explicitly verified that these solutions satisfy the equations
of motion as well as the following  Virasoro constraints
\begin{equation}
 \label{virspace}
 -\partial_{\pm}(u'+x')\partial_{\pm}(u'-x')+
\partial_{\pm}(y'+v')\partial_{\pm}(y'-v')=J^2.
\end{equation}
Note that compared to the Virasoro constraints of
solutions obtained by dressing time like geodesics in (\ref{virtime}) we have
$J^2 \rightarrow -J^2$.  This is expected  from the general proof that
the Virasoro constraints of the original solution
are preserved by the dressing method.
Since these are not  the  Virasoro constraints  obeyed by a physical string
these solutions  are best thought of as minimal surfaces in the  BTZ background.

\subsection{Closed string solutions}

The solutions given  in the equations (\ref{u1s}) are in general open strings  and
are complicated.
To get more insight into them we discuss the interesting situation
when  the solutions given in (\ref{u1s}) become periodic in $\sigma$.
Two special cases for the above solutions are $r=1$ and $\theta=\pi/2$. Note that for either of these two cases, $\alpha$ in (\ref{alp}) loses its $\sigma$ dependence.
For these cases the  solution given in  (\ref{u1s}) depends on $\sigma$ through the
trigonometric functions of the angle $\beta$ which in turn depends linearly on
$\sigma$.
Thus the solutions are periodic in $\sigma$ with the period determined by the coefficient
multiplying $\sigma$ in $\beta$ given in (\ref{alp}).
Thus the periodicity $\sigma_p$   of the closed string is given by
\begin{equation}
 \sigma_p = \frac{\pi(1+r^4-2r^2\cos(2\theta))}{Jr\sin(\theta)(1+r^2)}.
\end{equation}

\vspace{.5cm} \noindent
{\bf Closed strings with $r=1$, arbitrary $\theta$}
\vspace{.5cm}

We will discuss the case $r=1$ with arbitrary $\theta$ in some detail.
These solutions have $\sigma$ dependance through the periodic trigonometric functions. The solutions are now
closed strings   with period
\begin{equation}
 \sigma_p = \frac{2\pi}{J} \sin\theta.
\end{equation}
We will now evaluate the global charges for these solutions.
From the same analysis leading up to equation (\ref{rcharge}) we obtain the following
expression for  the charge $E+S$
\be\ba \label{scg1}
\frac{ 2( E+ S)}{\hat \lambda( r_+ - r_-)}
&= \int_{0}^{2\pi \sigma_p  } d\sigma {\rm Tr} ( \partial_0 \hat g \hat g^{-1} \sigma^1)
 - ( \lambda - \bar \lambda) {\rm Tr} (  (  P(\tau, 2\pi\sigma_p)  - P (\tau, 0 ) ) \sigma^1 ),
\\
&=  \int_{0}^{2\pi \sigma_p  } d\sigma {\rm Tr} ( \partial_0  g g^{-1} \sigma^3),
\\
&= \frac{ 4\pi (c_1 - c_2) \sin\theta}{J}.
\ea\ee
To obtain the second line we have used the periodicity property of the solution
and also the relation between the $SL(2, R)$ group element $g$ and the
$SU(1, 1)$ group element denoted by $\hat g$.
We have also verified the last line of the equation by explicitly evaluating
this charge on the dressed solution $g'$ given in (\ref{u1s}).
Thus we obtain the simple result that
the charge $E+S$ of the dressed solution reduces to the charge density of the
seed geodesic multiplied by the periodicity  of the string.
For the charge $E-S$ we obtain a similar result though we do not have a simple proof.
Explicitly evaluating the charge density $E-S$ for the solution (\ref{u1s})
we obtain
 \begin{eqnarray}\label{scg2}
\frac{ 2( E- S)}{\hat \lambda( r_+ - r_-)} = \frac{ 4\pi (c_1 + c_2) \sin\theta}{J}.
\end{eqnarray}
Here again we find that the charge is equal to the charge density
corresponding to the seed geodesic multiplied by the length of the closed string \footnote
{We have also verified that the property that the global charges are equal to the
charged density of the corresponding seed geodesic multiplied by the length
of the closed string remains true for the situation with $\theta = \frac{\pi}{2}$ and
arbitrary $r$. }.
Again the  equations for the charge in (\ref{scg1})  and (\ref{scg2}) closely resemble that
of giant magnon dispersion relations. Note that in this analysis we have chosen $\sin\theta >0$, whereas
for $\sin\theta<0$ the expressions in (\ref{scg1}) and (\ref{scg2}) are similar with
$\sin\theta$ replaced by $|\sin\theta|$.

We now qualitatively describe the motion of the string in the BTZ background.
From the solution given in (\ref{u1s}) we see that the coordinates
$u', v', x', y'$ assume large values when
\begin{equation}
\cot 2\beta^{*} = e^{\tilde \gamma},
\end{equation}
which implies
\begin{equation} \label{sbet}
\beta^{*}_{\pm}  =   \cot^{-1} ( e^{\tilde\gamma} \pm \sqrt{e^{2\tilde\gamma} +1} ).
\end{equation}
From (\ref{alp}) we see that for $r=1$ ,  $\beta$ is given by
\begin{equation}
\beta = \frac{J}{2} \left( \frac{\tau\cos\theta +\sigma}{\sin\theta} \right).
\end{equation}
Thus given $\tilde\gamma$,
 the equation (\ref{sbet}) determines a line in the worldsheet coordinates at which
 the space time coordinates become large.
 From the relationship between the radial coordinate of the BTZ to the
 coordinates $u', v', x', y'$ given in (\ref{reluvrt}) we see that  when
 the equation (\ref{sbet}) holds it is possible that the minimal surface can
 reach the boundary of  BTZ.
 
\begin{figure}
\centering
\includegraphics[width=70mm,height=50mm]{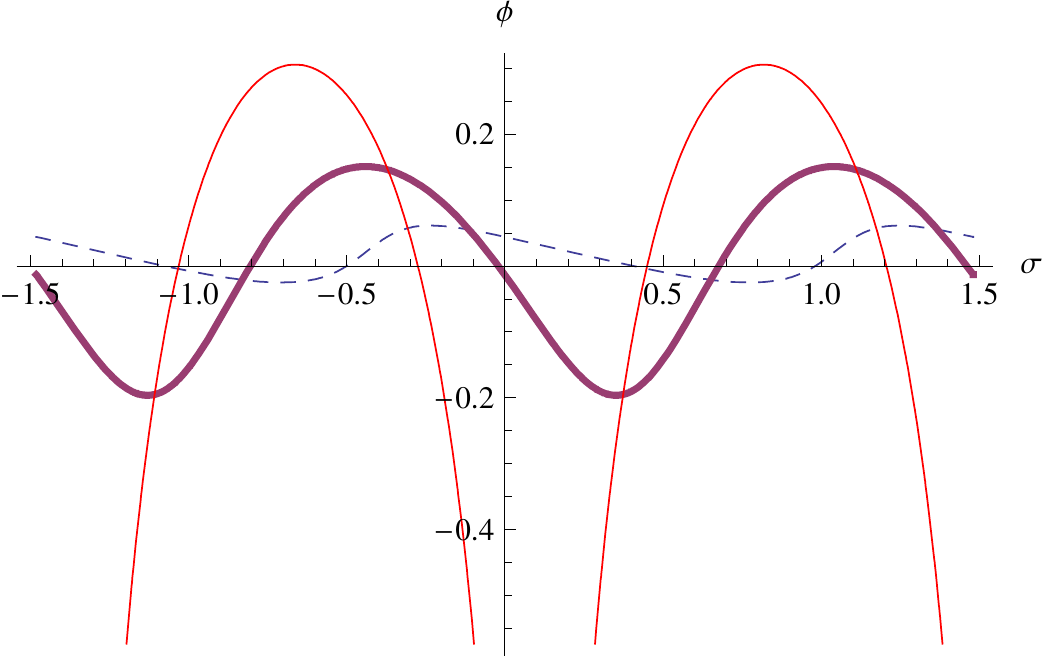}
\includegraphics[width=70mm,height=50mm]{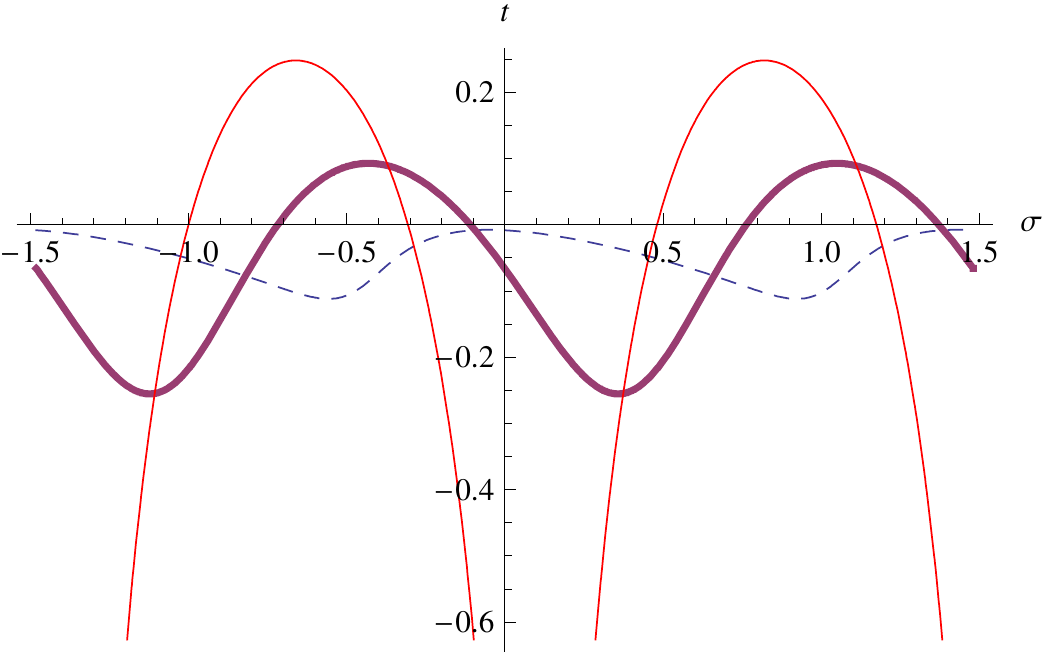}\\
\includegraphics[width=120mm]{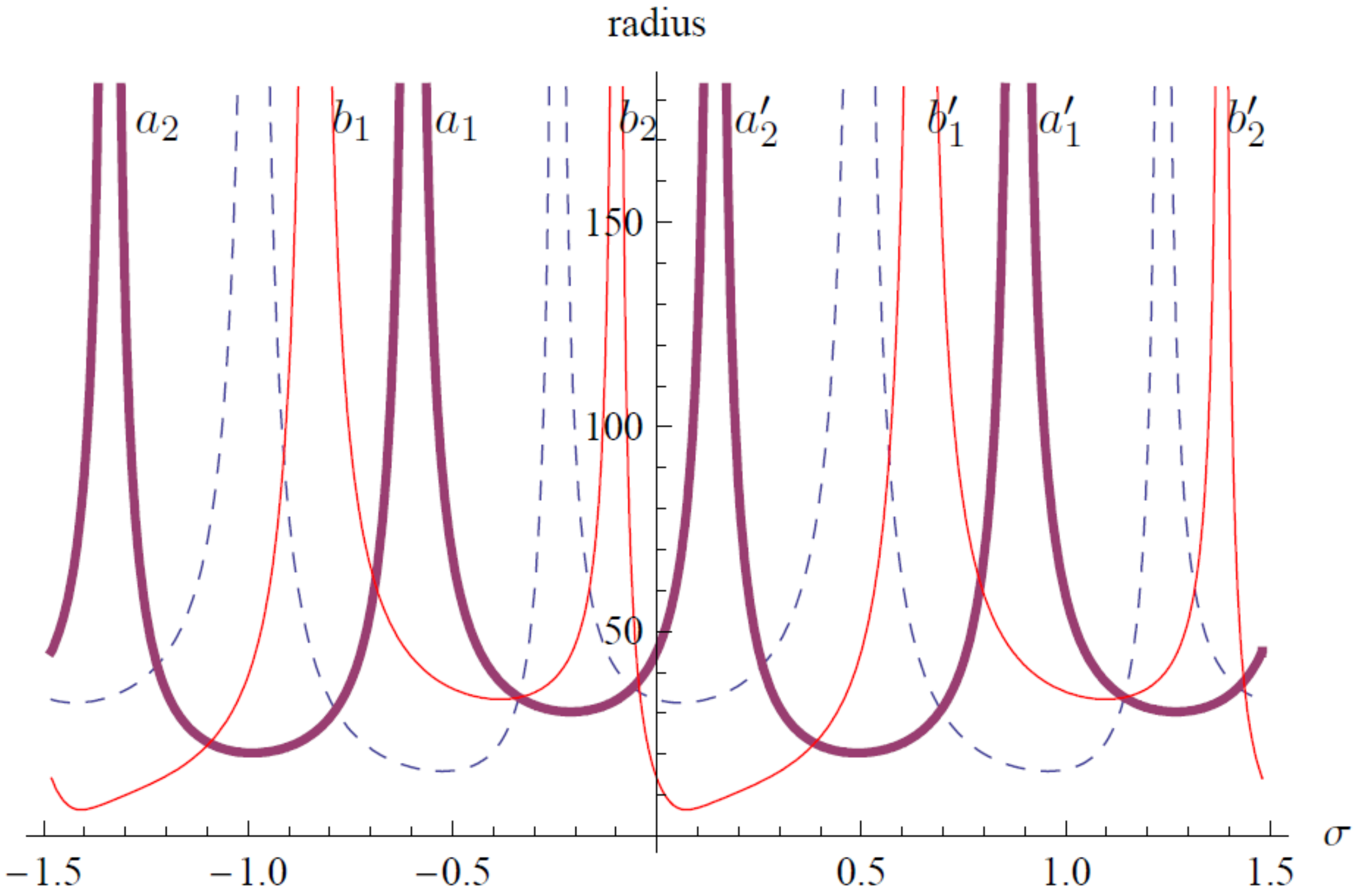}
\caption{Dashed ($\tau$=0.5), thick ($\tau$=1.0), red ($\tau$=1.34).}
\label{fig3}
\end{figure}

\begin{figure}
\centering
\includegraphics[width=100mm,height=70mm]{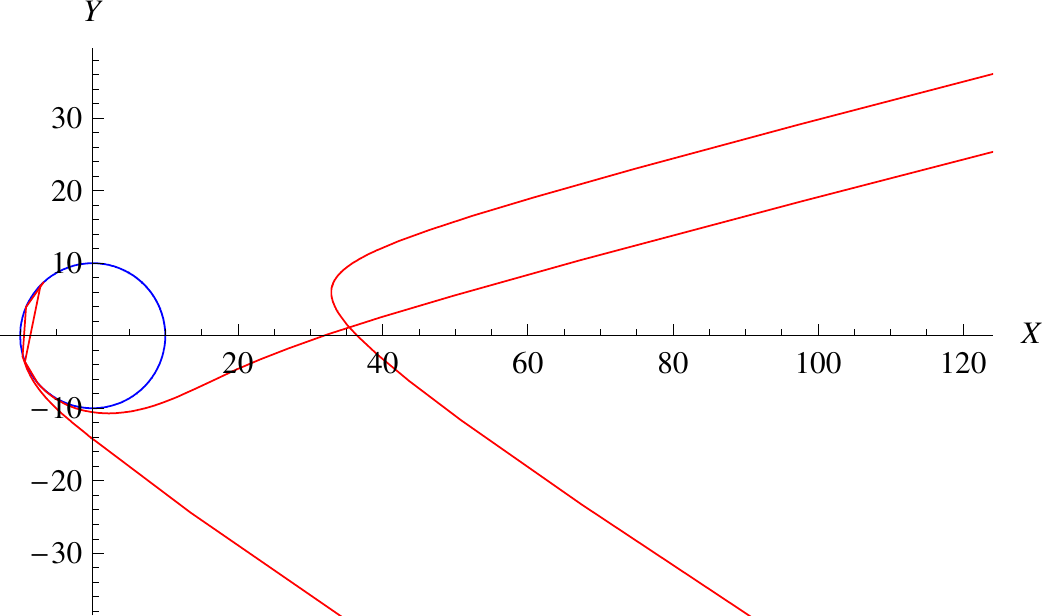}
\caption{Spacelike strings in $r-\phi$ plane at $\tau=1.34$.}
\label{fig4}
\end{figure}

\begin{figure}
\centering
\includegraphics[width=100mm,height=70mm]{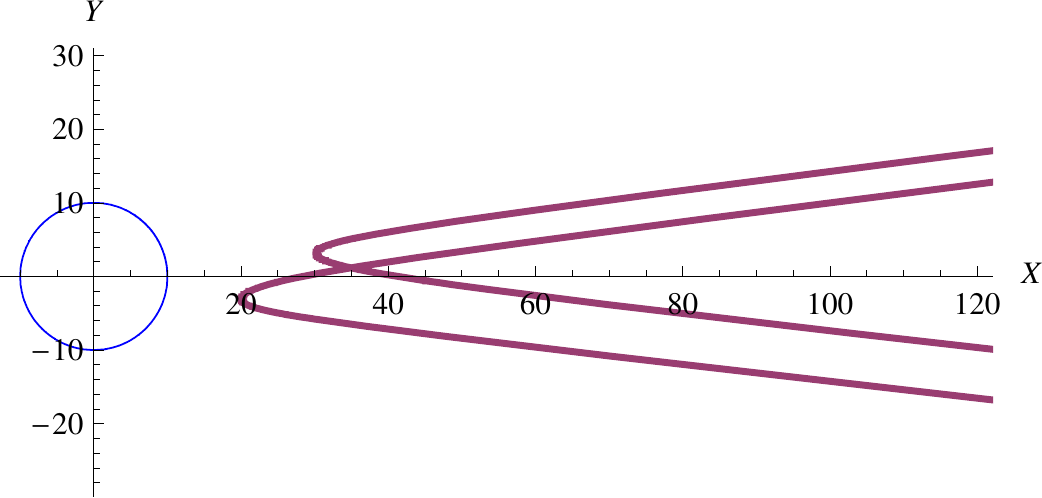}
\caption{Spacelike strings in $r-\phi$ plane at $\tau=1.0$.}
\label{fig5}
\end{figure}

\begin{figure}
\centering
\includegraphics[width=100mm,height=70mm]{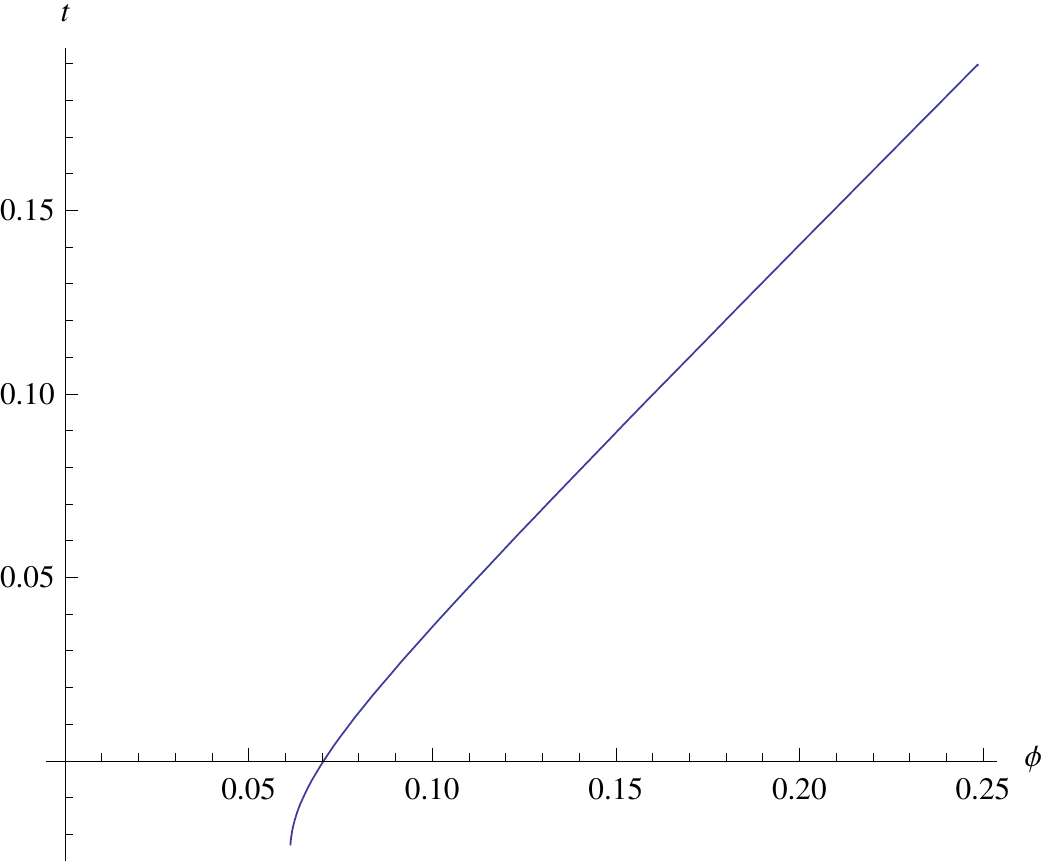}
\caption{Propagation of the BTZ boundary point due to root $\beta_1$ from $\tau=$0.5 to 1.34.}
\label{boundary1}
\end{figure}

\begin{figure}
\centering
\includegraphics[width=100mm,height=70mm]{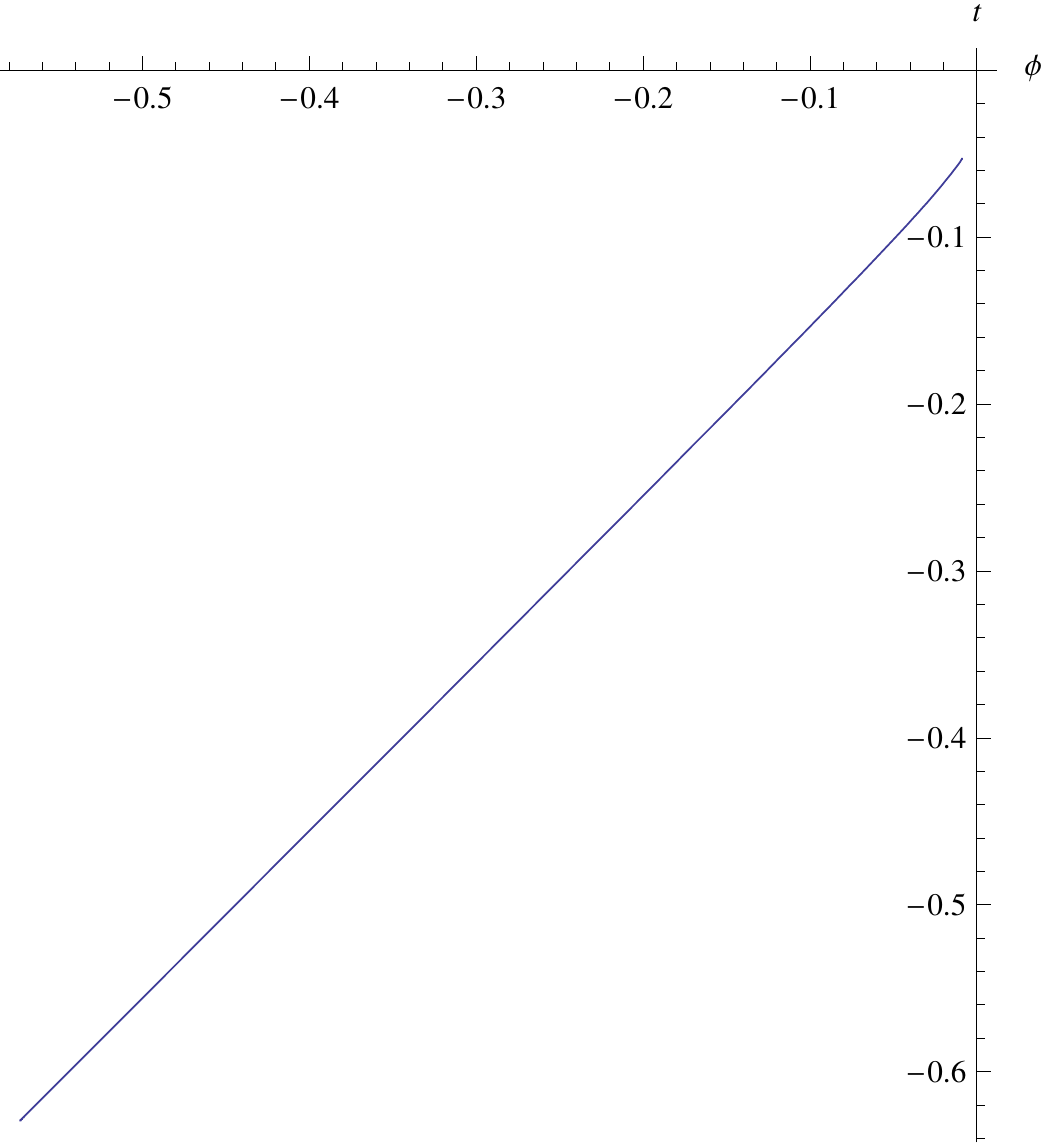}
\caption{Propagation of the BTZ boundary point due to root $\beta_2$ from $\tau=$0.5 to 1.34.}
\label{boundary2}
\end{figure}

 To describe the solution further we examine the solution at the following
 parameters
 \be
\label{par1}
\gamma_0=3,\quad r_+=10,\quad r_-=9,\quad J=3,\quad c_1=5,\quad c_2=1,\quad \theta=\frac{\pi}{4},\quad r=1.
\ee
As before the outer and inner horizon radius of BTZ is chosen to be $r_+= 10$ and
$r_- =9$ respectively.
Note that as discussed earlier and from figure \ref{hump} and figure \ref{figeo} the seed
space like geodesic for these
set of parameters originates close to the boundary and bounces back to the boundary
after reaching near the horizon. It does not penetrate the horizon.
The periodicity of the string for this set of parameters is given by
\begin{equation}
 \sigma_p = \frac{\sqrt{2} \pi}{3}.
\end{equation}
The values of $\beta^*_\pm$ are given by
\begin{equation}
 \beta^*_+ = 0.232484, \qquad
\beta^*_- =-1.33831.
\end{equation}
The curve on the world  sheet for which the space time coordinates become large is given by
\begin{equation}\label{relsigp}
 \sigma^*_\pm = \frac{\sqrt{2} }{3} \beta^*_\pm - \frac{\tau}{\sqrt{2}}.
\end{equation}
Thus given a value of $\tau$, there are two values of $\sigma$ at which
the space time coordinates are large.
To visualize the minimal surface we  plot its snapshot
at various values of worldsheet time.
 The radial $r$, time  $t$ and
the angular coordinate $\phi$ are plotted with respect to the $\sigma$ coordinate
at worldsheet times  $\tau= .5,~ 1.0,~ 1.34$ in figure \ref{fig3}.
The cross section of the minimal surface in the $r-\phi$ plane is provided
in figure \ref{fig4}  and figure \ref{fig5}.
The range of the $\sigma$ axis runs from $-\sqrt 2 \pi/3$ to $\sqrt 2 \pi /3$
in figure \ref{fig3}. The fact that the string is closed with period $\sqrt 2 \pi/3$ is clearly
visible in figure \ref{fig3}.  In figure \ref{fig3} we have also marked the position at which
the radial coordinate becomes infinite for the time $\tau =1.0,\, 1.34$.
For $\tau=1.0$, the value of $\sigma^*_+$ and its periodic image
is marked by $a_1$ and $a_1'$ and the value of $\sigma^*_-$  and its
periodic image is marked by $a_2$ and $a_2'$.
Similarly for $\tau=1.34$ the value of $\sigma^*_+$ and its image is marked
by $b_1$ and $b_1'$, the value of $\sigma^*_-$ and its image is marked by
$b_2$ and $b_2'$.
From the plots we see that the surface is pinned at the boundary on two curves
and hangs into the bulk. As the worldsheet time is increased
the surface crosses the horizon while being pinned at the boundary.
Eventually the string reaches the singularity.
We can also describe the two curves on the boundary at which
the minimal surface is pinned. These curves are given by
substituting the relations for $\sigma_+, \sigma_-$  given in (\ref{relsigp})  into the
expression for the time coordinate $t$ and the angular coordinate $\phi$
given in (\ref{reluvrt}). Then one can obtain a parametric plot of $t$ and $\phi$ with
respect to  the parameter $t$. These two curves corresponding to
$\beta_+^*$ and $\beta_-^*$  are given in  figure \ref{boundary1} and
figure \ref{boundary2} respectively.
From the figures we see that the curves describe a time like on the boundary.

\begin{figure}
\centering
\includegraphics[width=100mm,height=70mm]{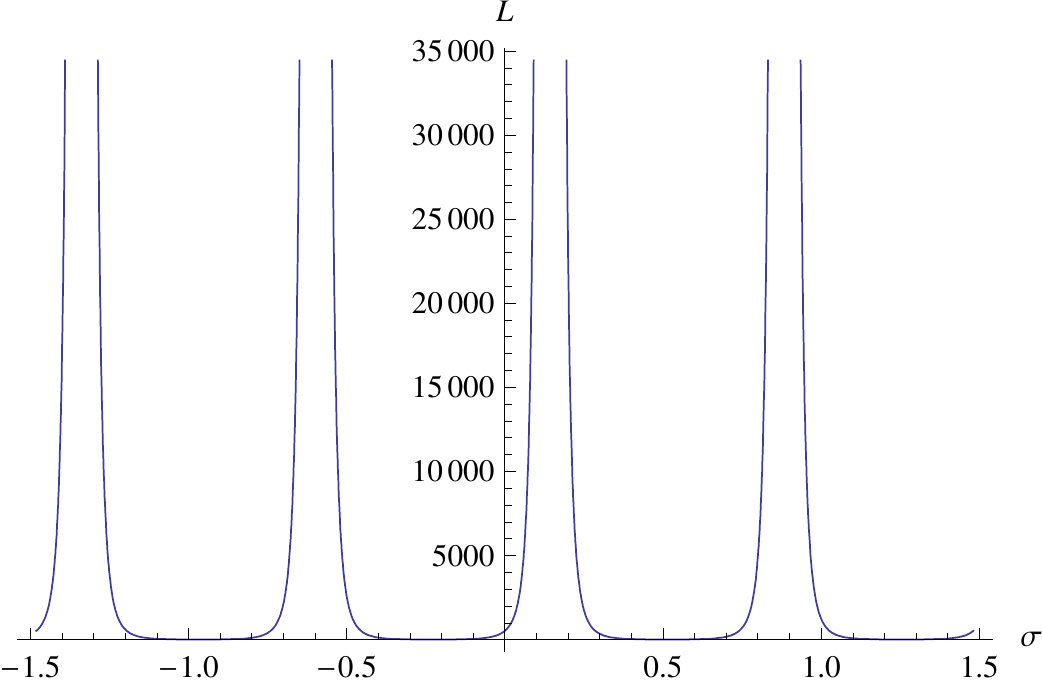}
\caption{Worldsheet Lagrangian vs $\sigma$ plot for space like strings.}
\label{nambu}
\end{figure}

Since these solutions can be interpreted
as   minimal surfaces it is interesting to verify if their proper area given
by the Nambu Goto Lagrangian is positive for all $\sigma$ for the
worldsheet times of interest \footnote{This check was suggested to us by Aninda Sinha.}.
The Nambu-Goto Lagrangian density
 is  given by
\be
\mathcal{L} _{NG}=\left[  {\rm Tr} ( g^{-1} g' g^{-1} \dot g ) )^2 -
{\rm Tr} ( g^{-1} \dot g g^- \dot g ) {\rm Tr} ( g^{-1} g' g^{-1} g') \right]^{1/2},
\ee
where $g$ is given in terms of  the  embedding  in (\ref{sl2g}).
Here the superscripts  prime and dot refer to derivatives with respect to
$\sigma$ and worldsheet time $\tau$ respectively.
 We have verified that this Lagrangian density is positive for all $\tau =0.5$ to $\tau=1.34$.
For a sample of its profile we have plotted the Lagrangian density
for $\tau = 1.0$ with respect to the worldsheet $\sigma$ coordinate in figure \ref{nambu}.

\newpage
\section{Giant gluons in BTZ}

The giant gluon solution found in \cite{Kruczenski:2002fb,Alday:2007hr}
 is a solution  of the Euclidean worldsheet
equations of motion
in $AdS_3$.
This solution has played an important role in evaluating the
gluon scattering amplitudes in $AdS$. It is interesting to study its properties in
the BTZ background. The embedding can be done in BTZ in two different ways consistent
with the periodicities of the BTZ background.

\vspace{.5cm}
\noindent
{\bf Method 1}
\vspace{.5cm}

In this method the giant gluon solution  can be written in terms of BTZ
embedding coordinates given in (\ref{sl2g}) as follows
\be\ba
 u&=\cosh \sigma \cosh \tau, \qquad &v&=\sinh \sigma \sinh \tau,  \\
x&=\sinh \sigma \cosh \tau , \qquad &y&=\cosh \sigma \sinh \tau.
\ea\ee
With the above choice the $SL(2,R)$ group element becomes
\be
 g= \left(
 \begin{array}{cc}
   \cosh \tau e^\sigma & \sinh \tau e^\sigma \\
  \sinh \tau e^{-\sigma} & \cosh \tau e^{-\sigma}
 \end{array}
 \right).
\ee
The conserved left and right currents are
\be\ba
j_\tau &=\partial_\tau g g^{-1}= \left(
 \begin{array}{cc}
   0 & e^{2\sigma} \\
  e^{-2\sigma} & 0
 \end{array}
 \right) , \qquad
&j_\sigma&=\partial_\sigma g g^{-1} = \left(
 \begin{array}{cc}
   1 & 0 \\
  0 & -1
 \end{array}
 \right), \\
l_\tau &= g^{-1}\partial_\tau g = \left(
 \begin{array}{cc}
   0 & 1 \\
  1 & 0
 \end{array}
 \right), \qquad
&l_\sigma&=g^{-1}\partial_\sigma g = \left(
 \begin{array}{cc}
   \cosh2\tau & \sinh2\tau \\
  -\sinh2\tau & -\cosh2\tau
 \end{array}
 \right).
\ea\ee
From these currents it can be seen that this solution satisfies the
 Euclidean worldsheet equations of motion trivially, i.e.
\begin{equation}
 \partial_\tau j_\tau + \partial_\sigma j_\sigma=0.
\end{equation}
They also satisfy the  Euclidean Virasoro constraints which are
\be\label{vire1}
-\partial_{\pm}(u+x)\partial_{\pm}(u-x)+\partial_{\pm}(y+v)\partial_{\pm}(y-v)=0,
\ee
where
$\partial_{\pm}=i\partial_{\tau} \pm \partial_{\sigma}$. The factor of $i$ in front of $\partial_\tau$ accounts for the fact that these solutions are
Euclidean.
The global charges $E$ and $S$ are zero for this configuration, since ${\rm{Tr}}(j_\tau \sigma_3)$ and ${\rm{Tr}}(l_\tau \sigma_3)$
vanish.

\begin{figure}
\centering
\includegraphics[width=100mm,height=70mm]{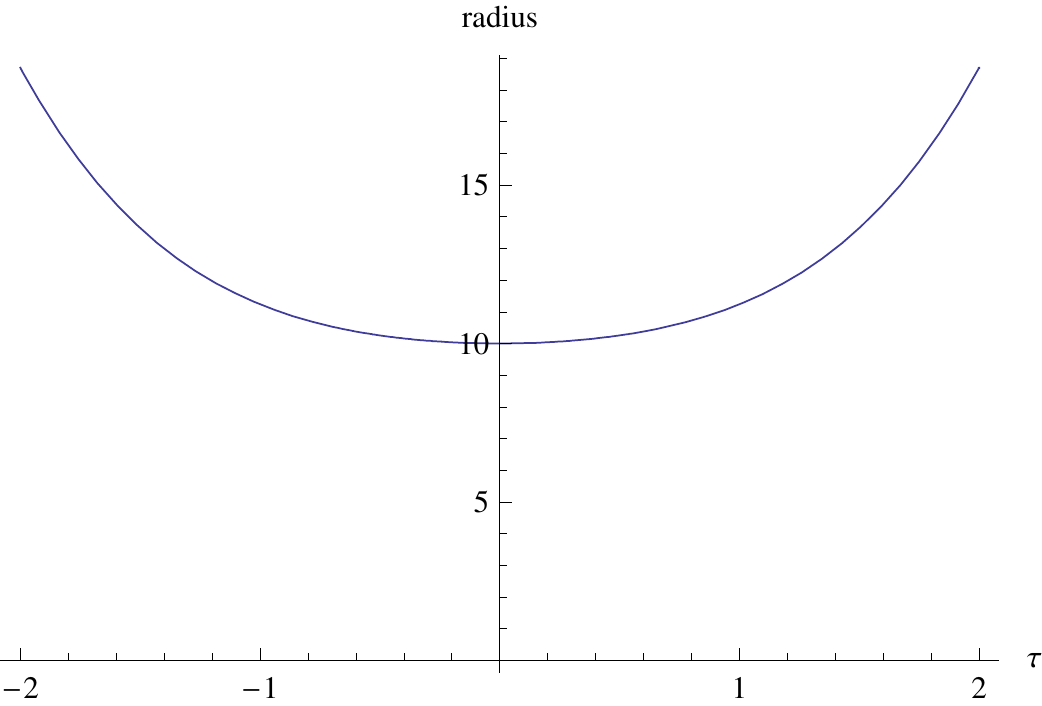}
\caption{Plot of $r$ with respect to $\tau$.}
\label{glur1}
\end{figure}

\begin{figure}
\centering
\includegraphics[width=70mm,height=60mm]{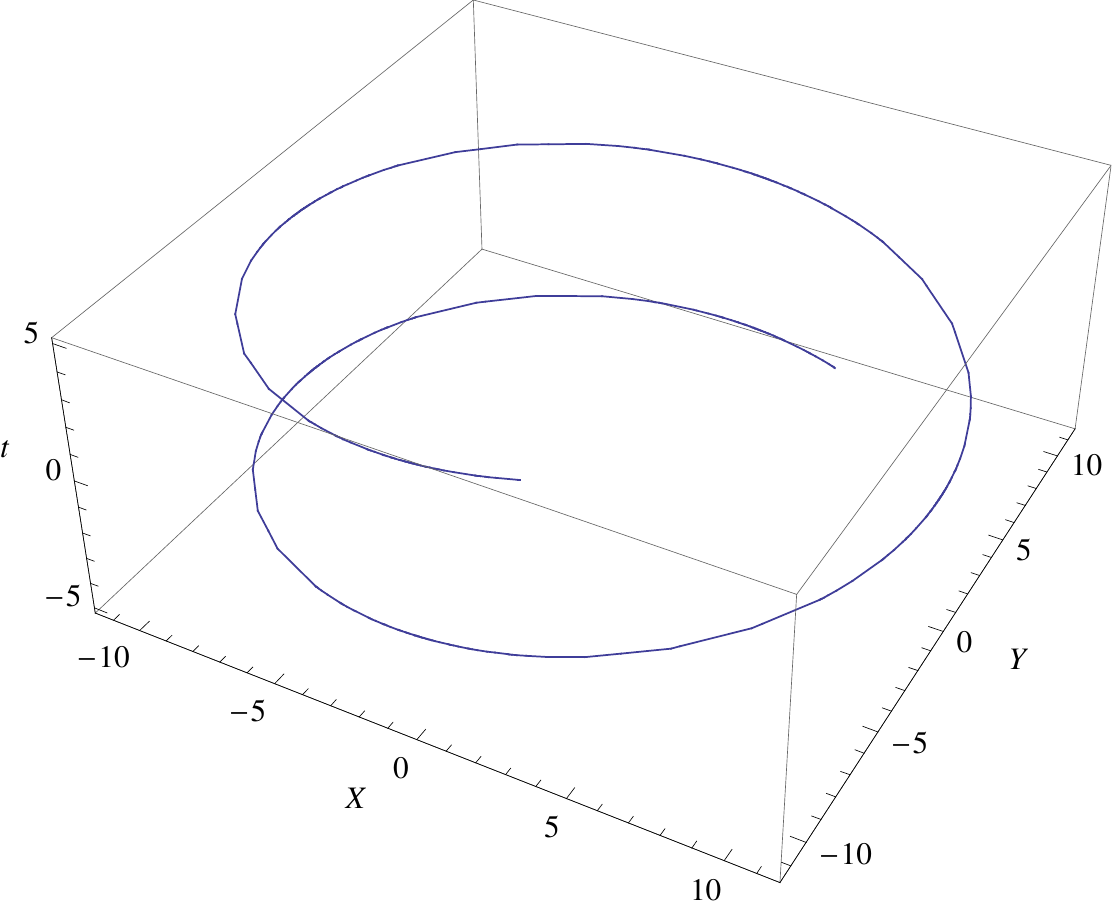}
\includegraphics[width=70mm,height=60mm]{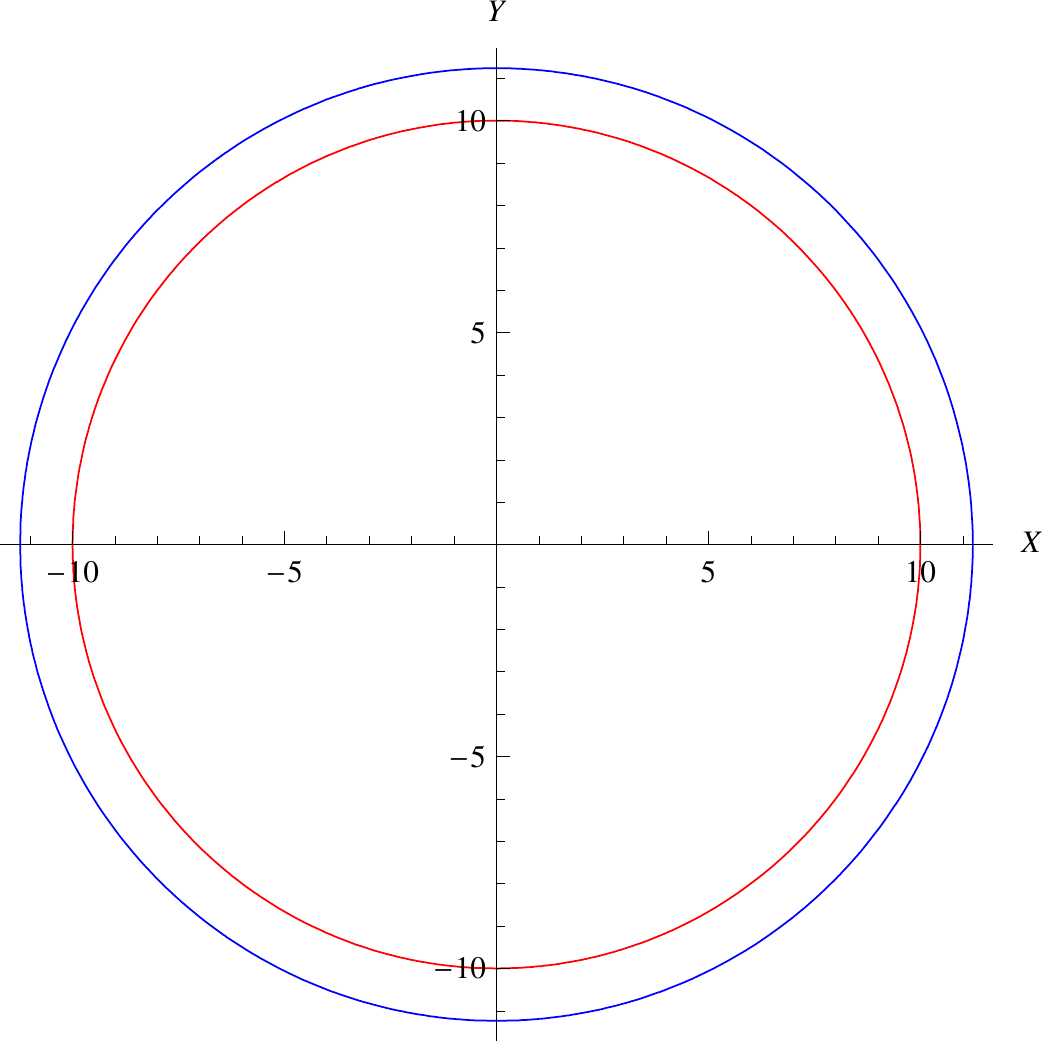}
\caption{The 3d plot shows a spiral string at $\tau=1$ and for $\sigma\in [-5,5]$ and the 2d plot shows the projection of the spiral in $r-\phi$ plane. The red circle denotes the position of the horizon.}
\label{glu3d1}
\end{figure}

Let us examine the behaviour of the solution in space time.
In terms of the BTZ coordinates  the solution has a rather simple form
\be\ba
r&=\sqrt{(r_+^2-r_-^2)(y^2-v^2)+r_+^2}=\sqrt{(r_+^2-r_-^2)(\sinh^2\tau)+r_+^2} ,\\
\phi&=\frac{r_+\log \left[\frac{u+x}{u-x}\right]+r_-\log\left[\frac{y+v}{y-v}\right]}{2(r_+^2-r_-^2)}=\frac{\sigma}{r_+-r_-},  \\
t&=\frac{r_-\log \left[\frac{u+x}{u-x}\right]+r_+\log\left[\frac{y+v}{y-v}\right]}{2(r_+^2-r_-^2)}=\frac{\sigma}{r_+-r_-}.
\ea\ee
 From figure \ref{glur1} we notice that $r$ is an even function of $\tau$ only and  the
 minimum of $r$ occurs at $r_+$ at $\tau=0$.  The space time coordinates
$t$ and $\phi$ are linear in $\sigma$. Thus  a snapshot of  the string
at a given value of $\tau$ has  a spiral shape as shown in figure \ref{glu3d1}.
As $\tau$ increases or decreases from
$\tau=0$, the radius of the spiral increases since $r$ is an even function of $\tau$.
Since this embedding has trivial global charges $E,~ S$ it is degenerate with the
BTZ state in the  CFT at least with respect to these charges.

\vspace{.5cm}\noindent
{ \bf Method 2}
\vspace{.5cm}

The other possible configuration for the string in $BTZ$ is written by interchanging $x$ and $y$ in
method 1.  This is given by
\be\ba
 u&=\cosh \sigma \cosh \tau , \qquad &v&=\sinh \sigma \sinh \tau ,\\
y&=\sinh \sigma \cosh \tau , \qquad &x&=\cosh \sigma \sinh \tau ,\\
\ea\ee
and the $SL(2,R)$ group  reduces to
\be
 g= \left(
 \begin{array}{cc}
   \cosh \sigma e^\tau & \sinh \sigma e^\tau \\
  \sinh \sigma e^{-\tau} & \cosh \sigma e^{-\tau}
 \end{array}
 \right).
\ee
Note that this embedding is obtained from the one in method 1 by the interchange of
$\sigma$ and $\tau$.
The conserved left and right currents are
\be\ba
j_\tau=\partial_\tau g g^{-1}= \left(
 \begin{array}{cc}
   1 & 0 \\
  0 & -1
 \end{array}
 \right) , &\qquad
j_\sigma=\partial_\sigma g = \left(
 \begin{array}{cc}
   0 & e^{2\tau} \\
  e^{-2\tau}& 0
 \end{array}
 \right), \\
l_\tau=g^{-1}\partial_\tau g = \left(
 \begin{array}{cc}
   \cosh2\sigma & \sinh2\sigma \\
 \sinh2\sigma & -\cosh2\sigma
 \end{array}
 \right),  &\qquad
l_\sigma=g^{-1}\partial_\sigma g = \left(
 \begin{array}{cc}
   \cosh2\tau & \sinh2\tau \\
  -\sinh2\tau & -\cosh2\tau
 \end{array}
 \right).
\ea\ee
This solution satisfies  and Euclidean equations of motion and the
corresponding  Virasoro constraint  given in (\ref{vire1}).
Unlike the case of the embedding in method 1,
this solution has non zero values for $E$ and $S$.  It is
\be\ba
 E+S&=\frac{\hat \lambda}{2}(r_+-r_-)\int_{-\Lambda}^{\Lambda}d\sigma {\rm{Tr}}(j_\tau \sigma_3)  \\
&= 2\Lambda\hat \lambda(r_+-r_-),
\ea\ee
and
\be\ba
 E-S&=-\frac{\hat\lambda}{2}(r_++r_-)\int_{-\Lambda}^{\Lambda}d\sigma {\rm{Tr}}(l_\tau \sigma_3) \\
 &= -\frac{\hat\lambda}{2}(r_++r_-)e^{2\Lambda},
\ea\ee
where we have introduced the cut-off  $\Lambda$.
For large $\Lambda$ we can solve for $S$ and eliminate the cut-off to obtain
the dispersion relation
\begin{equation}
E+ S \sim \hat\lambda (r_+ - r_-) \ln \left( \frac{4S}{\hat \lambda( r_+ + r_-) } \right).
\end{equation}
These dispersion relations are similar to that
seen in the giant gluon solution.

\begin{figure}
\centering
\includegraphics[width=90mm,height=50mm]{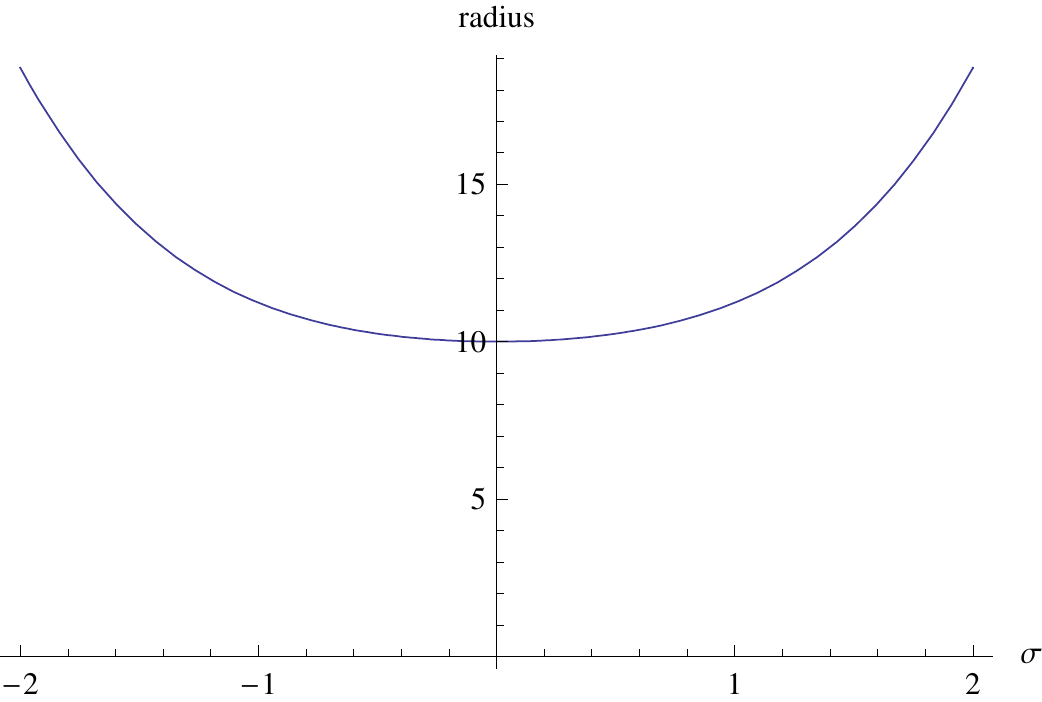}
\caption{Variation of $r$ with respect to $\sigma$.}
\label{glur2}
\end{figure}

\begin{figure}
\centering
\includegraphics[width=70mm,height=80mm]{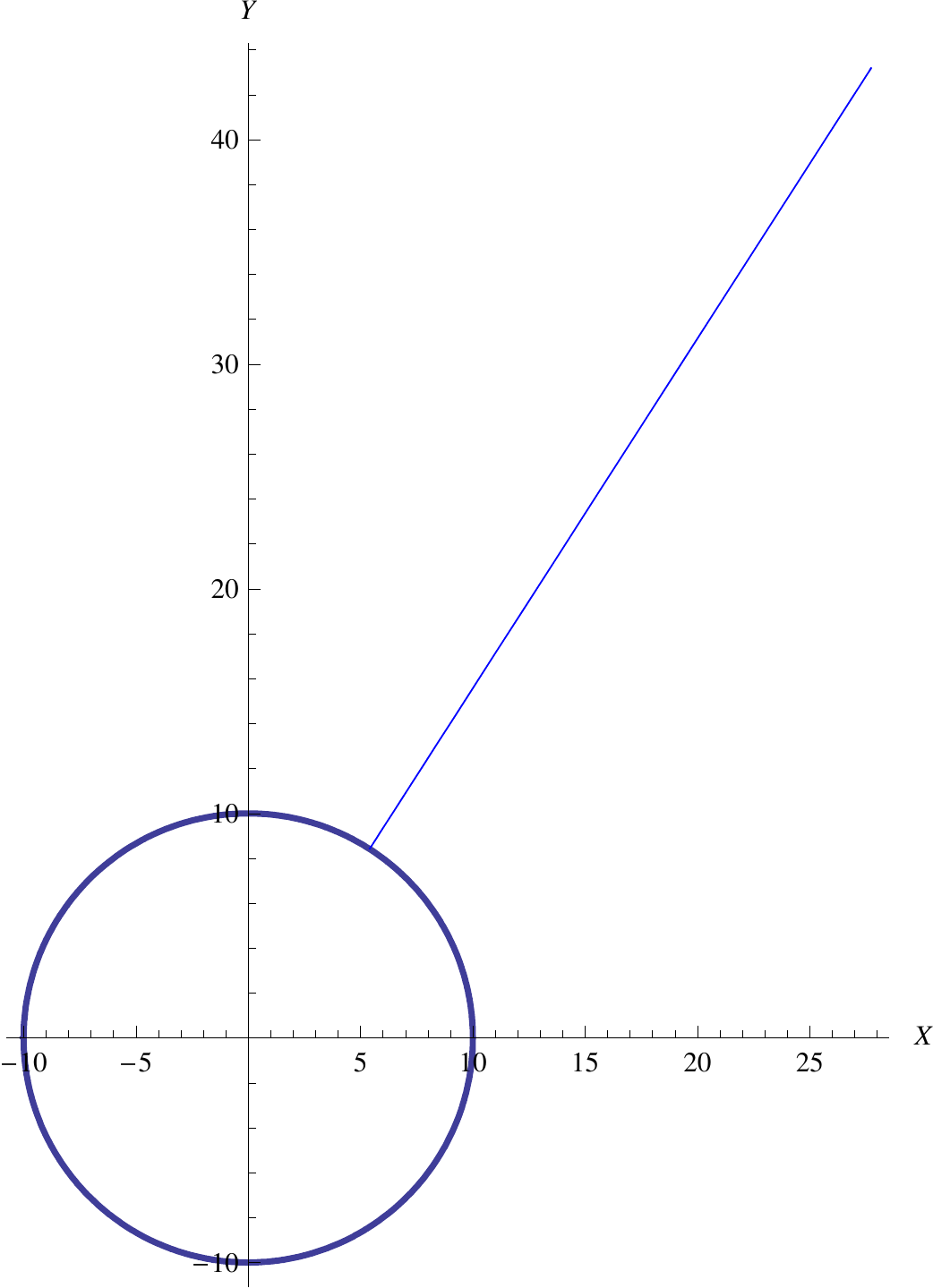}
\includegraphics[width=70mm,height=80mm]{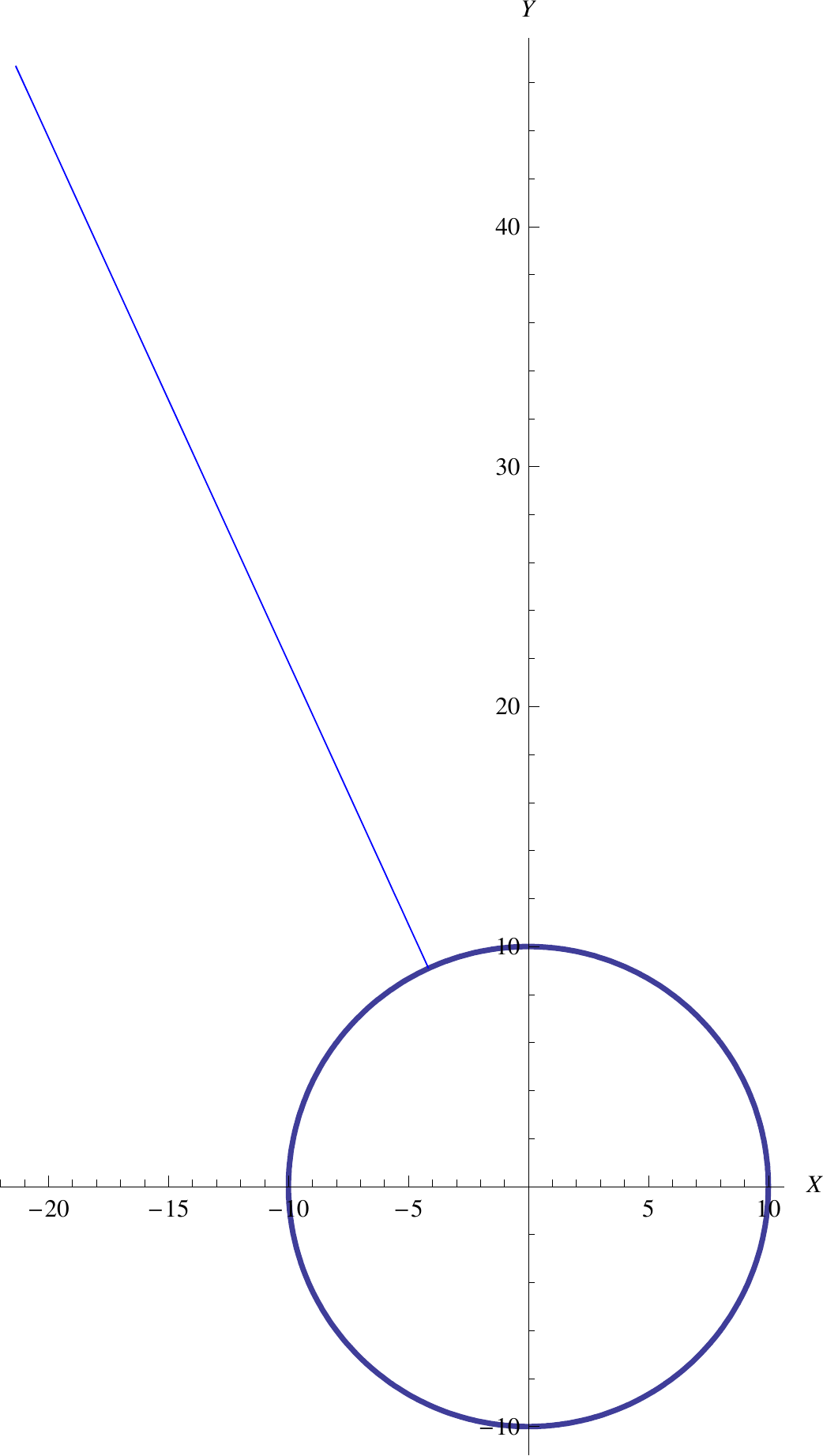}
\caption{The position of string in $r-\phi$ plane at $\tau=1$ (left) and $\tau=2$ (right) plotted for
$\sigma\in[-\pi,\pi],~r_+=10$ (blue circle), $r_-=9$.}
\label{gluxy}
\end{figure}

The space time  coordinates for this embedding is given by
\be
r=\sqrt{(r_+^2-r_-^2)(\sinh^2\sigma)+r_+^2}, \qquad \phi=t=\frac{\tau}{r_+-r_-},
\ee
where $r$ is again an even function of $\sigma$ and it's minimum occurs at $r_+$ as seen in figure \ref{glur2}. Figure \ref{gluxy}  shows  the snapshot  of the string at two different $\tau$'s.
The open string is a spike which originates at the boundary, touches the horizon and returns back to the
boundary.  It is folded onto itself.
Since $\phi$ is linear
 in $\tau$,  the string will rotate around the horizon as $\tau$ changes. The end points of the string
 at the boundary describe a light like trajectory.

\section{Conclusions}

We have applied the dressing method to construct and study classical string solutions
in the BTZ background. Dressing time like and space like geodesics we obtained
solutions which have dispersion relation similar to that of the giant magnon.
Dressed time like geodesics are open strings whose end points move on time like
trajectories with one point  pinned to the boundary.  They eventually fall into the horizon.
We studied closed strings obtained by dressing space like geodesics. The minimal surfaces
obtained  are pinned on the boundary at two points and can cross the horizon.
Giant gluons can be also embedded in the BTZ background. They give rise to
two solutions, one of them is a spiral which expands and contracts, the other is a
spike stretching from the boundary to the horizon.

There are several interesting directions which can be pursued to  develop the observations
of this paper further.  It will be interesting to study the solutions obtained by
 multiple dressings and find their general properties, the appendix contains a discussion
of the  general procedure to obtain these solutions.
  If their dispersion
relations turns out to be  similar to multi-magnon solutions  then perhaps these
are the fundamental excitations of the integrable system corresponding to
the BTZ background.  Another direction is to investigate the role of these solutions
from the boundary CFT point of view.  The solution obtained by dressing space like
geodesic is possibly best suited to be thought of  as  a non-local
observable in the CFT. It will be interesting to pin down this observable precisely.
Since these solutions penetrate the horizon while being pinned to the boundary,
it will be extremely interesting to quantize the fluctuations about these solutions.
This will  help to verify the general  predictions of the behaviour of strings near the horizon
as argued in \cite{Susskind:1993aa} and give more insight of the physics of
extended objects near the horizon.

To summarize:  Classical string propagation in the BTZ sigma model is integrable.
 As shown in this paper it can be used to construct several interesting
classical solutions in this background.
This structure of the BTZ black hole is certainly worth  further investigation.

\acknowledgments

We thank George Jorjadze,
Chethan Krishnan, S. Prem Kumar, Gautam Mandal, Georgios Papathanasiou, Francisco Rojas and  Aninda Sinha
for useful comments and discussions.
C.K thanks CHEP, IISc for hospitality during which this work was initiated.
J.R.D  and C.K thank the organizers of the
``Recent Advances in Quantum Field Theory and String Theory'' for a stimulating  conference
held at Tbilisi in 2011 during which some parts of this work was done.
The work of  J.R.D is partially supported by
the Ramanujan fellowship DST-SR/S2/RJN-59/2009, the work of A.S is supported by  a
CSIR fellowship (File no: 09/079(2372)/2010-EMR-I).

\appendix
\section{$N$-dressed geodesics}
Once we know a solution $\Psi_0(\lambda)$ to the system \eqref{mon} it is possible to construct an explicit solution for the $N$-dressed geodesic.   The details of this calculation for the $SU(2)$ case have been worked out in \cite{Kalousios:2010ne}.  A parallel treatment for $SU(1,1)$ yields that the $N$-th dressed solution $g_N$ is
\be\label{g_N=}
g_N=\frac{\prod_{i=1}^N \frac{\l_i}{|\l_i|}}{{\rm det}(a_{ij})}
:\left|
\begin{matrix}
 g_0 & -h_1 & \cdots & -h_N \\
 h_1^\dagger M g_0 & a_{11} & \cdots & a_{1N} \\
 \vdots & \vdots & \vdots & \vdots \\
 h_N^\dagger M g_0 & a_{N1} & \cdots & a_{NN}
\end{matrix}
\right|: \, ,
\ee
where the colons around the second determinant simply mean that upon expanding the determinant, the column $h_i$ is ordered before the row $h_j^\dagger M g_0$.  In the above
\be
h_i=\Psi_0(\bar{\l}_i) e_i, \qquad \a_{ij}=-\frac{\l_i \beta_{ij}}{\l_i-\bar{\l}_j}, \qquad \beta_{ij}=h_i^{\dagger} M h_j.
\ee
The parameter $\l_i$ is the spectral parameter and $e_i$ the polarization vector.

As an example we give the first two dressed solutions
\be\ba
g_1&=\frac{\l_1}{|\l_1|}\left( 1+ \frac{ h_1 h_1^{\dagger} M}{\a_{11}} \right)g_0,\\
g_2&=\frac{\l_1 \l_2}{|\l_1 \l_2|}\left(1+ \frac{\a_{22}h_1 h_1^{\dagger} M +\a_{11}h_2 h_2^{\dagger} M-\a_{12}h_1 h_2^{\dagger} M-\a_{21}h_2 h_1^{\dagger} M}{\a_{11}\a_{22}-\a_{12}\a_{21}} \right) g_0.
\ea\ee

In order to prove the above \eqref{g_N=} we start by applying consecutive dressing transformations to the vacuum solution $\Psi_0(\lambda)$.  Then, at the $N$-th step we will have $\Psi_N(\l)=\chi_N(\l)\Psi_{N-1}(\l)$, with a dressing factor
\be\label{chi_N}
\chi_N(\l)=1+\frac{\l_N-\bar{\l}_N}{\l-\l_N}{\cal P}_N.
\ee
In the above, ${\cal P}_N$ is a rank one projection operator that  can be expressed as
\be
{\cal P}_N=\frac{1}{\beta_{NN}^{(N-1)}}\; h_N^{(N-1)} h_N^{(N-1)\dagger}M,
\ee
where we have defined
\be
h_i^{(N)}=\Psi_N(\bar{\l}_i)e_i, \quad \beta_{ij}^{(N)}=h_i^{(N)\dagger}M h_j^{(N)},\quad \a_{ij}^{(N)}=-\frac{\l_i \beta_{ij}^{(N)}}{\l_i-\bar{\l_j}},
\ee
for any $N\ge0$. We can then obtain a recursion relation for the matrix field $g_N=\Psi_N(0)$ which will be given by
\be\label{gn}
g_N=\frac{1}{\alpha^{(N-1)}_{NN}}\big(\alpha^{(N-1)}_{NN}+h^{(N-1)}_N h^{(N-1)\dagger}_N M\big) g_{N-1}.
\ee
Furthermore, we can use the relations (\ref{chi_N})-(\ref{gn}), and obtain the following recursion relations
\be\ba\label{hgSU}
h_i^{(N)} &= h_i^{(N-1)}- \frac{\a_{Ni}^{(N-1)}}{\a_{NN}^{(N-1)}} h_N^{(N-1)} \\
\a_{ij}^{(N)} &=\a_{ij}^{(N-1)}-\frac{\a_{iN}^{(N-1)}\a_{Nj}^{(N-1)}}{ \a_{NN}^{(N-1)}}\\
h_i^{(N)\dagger}M g_N &= \frac{ 1}{\a_{NN}^{(N-1)}}\big(\a_{NN}^{(N-1)} h_i^{(N-1)\dagger}M g_{N-1} -\a_{iN}^{(N-1)} h_N^{(N-1)\dagger}M g_{N-1}\big),
\ea\ee
which after a straightforward calculation will lead to the final expression \eqref{g_N=}.

\bibliography{btz22}
\bibliographystyle{JHEP}

\end{document}